%Paper: astro-ph/9302001
%From: ethan@grendel.as.utexas.edu (Ethan Vishniac)
%Date: Wed, 3 Feb 93 10:31:06 CST

%Updated with effect from: 5 Sept 1991
\headline={\ifnum\pageno=1\firstheadline\else
\ifodd\pageno\rightheadline \else\leftheadline\fi\fi}
\def\firstheadline{\hfil}
\def\rightheadline{\hfil}
\def\leftheadline{\hfil}
        \footline={\ifnum\pageno=1\firstfootline\else\otherfootline\fi}
\def\firstfootline{\rm\hss\folio\hss}
\def\otherfootline{\hfil}
\font\tenbf=cmbx10
\font\tenrm=cmr10
\font\tenit=cmti10
\font\elevenbf=cmbx10 scaled\magstep 1
\font\elevenrm=cmr10 scaled\magstep 1
\font\elevenit=cmti10 scaled\magstep 1

%\TagsOnRight
\nopagenumbers
\parskip=0pt
\line{\hfil }
\vglue 1cm
\hsize=6.0truein
\vsize=8.5truein
\parindent=3pc
\baselineskip=10pt
\centerline{\tenbf ANGULAR MOMENTUM TRANSPORT IN LOW MASS ACCRETION DISKS}
\vglue 1.0cm
\centerline{\tenrm ETHAN T. VISHNIAC}
\baselineskip=13pt
\centerline{\tenit Department of Astronomy, University of Texas}
\baselineskip=12pt
\centerline{\tenit Austin, TX 78712, USA}
\vglue 0.3cm
\centerline{\tenrm and}
\vglue 0.3cm
\centerline{\tenrm PATRICK DIAMOND}
\centerline{\tenit Department of Physics, University of California, San Diego}
\baselineskip=12pt
\centerline{\tenit La Jolla, CA 92093, USA}
\vglue 0.8cm
\centerline{\tenrm ABSTRACT}
\vglue 0.3cm
{\rightskip=3pc
\leftskip=3pc
\tenrm\baselineskip=12pt%\parindent=1pc
\noindent
We consider the transport of angular momentum in accretion disks that are
low mass, in the sense that the gravitational forces produced by the
material in these disks has a negligible effect on disk dynamics.
There is no established consensus on how this transport takes place.
We note that for phenomenological reasons the traditional $\alpha$
model is probably not a good description of real disks.
Here we briefly review a few of the more promising models.
These include models in which angular momentum transport is driven
by shocks, and by magnetic field instabilities.  The latter
is more promising, but requires a dynamo.  We note that the
direction of angular momentum transport due to convection
in a conducting disk is not known as competing mechanisms are at
work.  We briefly discuss a number of
possible dynamo mechanisms, and their problems.
We then give a detailed exposition of the internal wave driven dynamo
model, in which internal waves excited at large radii drive
an $\alpha-\Omega$ dynamo.  The azimuthal magnetic field produced in this
way is unstable to a magnetic shearing instability (MSI), which drives nearly
isotropic turbulent eddys with typical fluid velocities of $\sim V_A$,
where $V_A$ is the local Alfv\'en speed.  The scale of the eddys is
$\sim V_A/\Omega$, where $\Omega$ is the local rotational frequency.
This turbulence leads to a saturation of the dynamo when
$V_A\sim (H/r)^{2/3} c_s$, where $H$ is the half-thickness of the
disk, $c_s$ is the local sound speed, and $r$ is the radial coordinate.
This gives rise to an effective dimensionless viscosity coefficient
$\sim (H/r)^{4/3}$ and vertical and radial diffusion coefficients
which are $\sim (H/r)^{4/3} Hc_s$.  The resulting vertical diffusion of entropy
will have a substantial effect on detailed models of vertical structure
in accretion disks.  Viscous and thermal instabilities of very hot
disks, those dominated by radiation pressure and electron scattering,
are substantially moderated in this model.  We note that the MSI largely
suppresses the Parker instability in accretion disks.
\vglue 0.8cm }
\line{\elevenbf 1. Introduction\hfil}
\vskip .4cm
\elevenrm\baselineskip=14pt

The study of astrophysical disks has, at its center, a curious and enduring
puzzle,
{\elevenit i.e.}
the radial transport of angular momentum.  Throughout this book
such transport is assumed to take place at a rate much higher than one would
obtain
by appealing to well understood microscopic processes.  The evident success of
such models, in spite of their ad hoc basis, suggests that some kind of
collective
process (or processes) is responsible.  Here we will describe recent work
aimed at producing a model of these processes.  What follows will not be
completely
objective.  There exists no well-defined consensus in this field and no model
is free from serious objections.  This paper describes what {\elevenit we}
think
are the most promising approaches to this problem.  Other workers will
have different opinions.  A popular, although somewhat dated, reference for
this topic
is Pringle 1981, who quite forthrightly declared that very little was known
about disk
viscosity at that time. We believe, perhaps optimistically, that times have
changed.
Nevertheless, much of the following discussion will be qualitative, which is
indicative
of the large uncertainties that remain.

We begin by defining the problem we wish to solve in the most narrow possible
terms.
First, we are concerned
with {\elevenit thin} accretion disks, {\elevenit i.e.}
those for which the disk thickness $H$
is much less than the radius $r$.  This is not because we consider thick disks
to be rare or uninteresting, but because $(H/r)$ is a natural ordering
parameter.
When it is large then the problem becomes significantly more complicated.
Second, we assume that the local rotational frequency, $\Omega(r)$, is
essentially
Keplerian, {\elevenit i.e.}
$\Omega^2=GM_*/r^3$.  If the disk is thin, then this is equivalent to assuming
that $M_{disk}\ll M_*$.  Third, we will assume that our disks have negligible
self-gravity.
In other words, their vertical gravity is approximately $-z\Omega(r)^2$, where
$z$ is the
distance to the midplane.  This implies that $M_{disk}\ll (H/r)M_*$.
A disk that violates this condition will suffer from local gravitational
instabilities which are outside the scope of this paper.  We note that as a
consequence of
these three conditions $c_s\sim H\Omega$, where $c_s$ is the sound speed at the
disk midplane.  Fourth, we ignore any externally imposed magnetic field.
Fifth, we assume that the disk is perfectly conducting.  These conditions
need not be met everywhere within a disk.
Rather we expect that we can confine our attention to
those regions within a disk where these conditions are met.  Viewed in this
light these
are not particularly onerous conditions and will apply to the majority of
visible systems.  The chief exceptions are protostellar disks, which are likely
to be poorly
conducting.  We will ignore their existence.

The traditional treatment of accretion disks relies on the introduction of
a phenomenological viscosity given by $\nu=\alpha c_sH$ (Shakura and Sunyaev
1973).
This expression was originally motivated by the expectation that the high
Reynolds number of accretion disk flows would lead to strong, approximately
isotropic,
turbulence.  Since $H$ and $c_s$ are the only natural scales for length and
velocity
this implies an $\alpha$ of order unity, the exact value of which is only
weakly
dependent on the specific nature of the disk in question.  Since then it
has become clear that there are no purely hydrodynamic linear instabilities in
accretion disk flow.  Nevertheless there still may be some justification
for this approach.  Processes which are purely {\elevenit local} in the sense
that
they do not rely on input from distant regions of the disk will tend to
reduce to this formalism since $H$ and $c_s$ remain the only natural scales.
This includes local MHD processes, since the only `natural' scale for the
magnetic
pressure is the pressure of the conducting fluid.  Quite apart from the
question
of whether or not $\alpha$ models will turn out to be justified due to some
local instability, they have been extremely useful in constructing
phenomenological
models of disks.  This is due, in part, to the fact that the description of a
stationary,
optically thick disk is independent of the precise nature of the disk
`viscosity'.
Still, real disks are not necessarily either stationary or optically thick and
it should be possible to constrain models of angular momentum transport by
appealing
to observational work.  In general, nonlocal models will not be equivalent to
$\alpha$ models, but
it is still convenient to quote results in terms of an equivalent ``$\alpha$''
for a stationary disk.

It is worth noting that it is not clear that a purely local theory that is
really
equivalent to an $\alpha$ model can be made to work at all.  A number of
attempts
have been made to model dwarf novae outbursts as the result
of a thermal instability in accretion disks around white dwarfs and neutron
stars
(Mineshige and Osaki, 1983; Meyer and Meyer-Hofmeister, 1984; Meyer, 1984;
Smak, 1984;
Mineshige and Osaki, 1985; Lin, Papaloizou, and Faulkner, 1985;
Cannizzo, Wheeler, and Polidan, 1986).  Similar work has also been done on
X-ray transients (Huang and Wheeler, 1989; Mineshige and Wheeler, 1989).
Such models can produce quite successful results, but only if the
value of $\alpha$ can be made to rise significantly as the disk temperature
goes up.  Moreover Wood and Mineshige (1989) have argued that during quiescence
the emission profiles of the disks of Z Cha and OY Car imply  that $\alpha$
decreases
at small radii as $r^{0.3-0.4}$.  These results can be summarized as suggesting
that $\alpha$
scales as $H/r$ to some power of order unity.  It is not clear how
a model based on local processes can produce this kind of scaling.

A useful way to look at angular momentum transport processes is to start by
examining the available linear MHD modes, and see what roles they might play.
We do this in section 2 of this paper, at the same time reviewing several
candidate processes.  In section 3 we will describe the basic theory
of magnetic dynamos in disks and discuss some previous attempts to
construct a self-consistent disk dynamo.
In section 4 we return to the focus of
our current work, the internal wave driven dynamo and show how to derive
the critical scaling arguments that describe its operation.  In the final
section we will discuss the prospects for progress in this contentious field.
\vskip .6cm
\line{\elevenbf 2. Local Modes\hfil}
\vskip .4cm
Our discussion of the local modes within a disk will be based on a number of
simplified situations, which will suffice to illustrate the critical
physical processes.  We begin by making a methodological point.  There are
a variety of equally valid ways to examine modes in a system which is
undergoing
some zeroth order motion.  One common method used in this context is to
describe
disturbances in terms of some radial plane wave, {\elevenit i.e.}
$f(r,\theta)\propto\exp[i(k_rr+m\theta)]$.
The vertical structure may be included explicitly, giving a nontrivial
dependence on
$z$, or, for some applications, its effects may be ignored and the perturbation
may
be assumed to be proportional to $\exp(ik_zz)$.  In either case the underlying
differential rotation of the disk will give the perturbation a complicated
dependence on time, including a linear increase in $k_r$ with time.
We will prefer to describe perturbations as rigidly rotating patterns, with a
simple dependence on time, but a complicated radial structure.  In general
we feel that this makes the evolution of waves and instabilities more
transparent.

We start by considering local adiabatic hydrodynamic disturbances in a disk.
In this case `local' means that the typical radial length scales associated
with the perturbations are much less than the radius of the disk.  We will
therefore neglect all radial derivatives of the background fluid properties
and the geometry.  There is one important radial derivative that survives
(and explains why the radial eigenfunctions are not pure plane waves).
This is because $\partial_t$ always appears in the combination
$\partial_t+\Omega\partial_\theta=\partial_t+im\Omega$.  This can be understood
physically as being due to the fact that the physically important frequency
is not the one measured by an observer in an inertial frame, but rather
the frequency of the perturbation measured by an observer moving with the
unperturbed fluid.  This is $\bar\omega\equiv\omega+m\Omega$.
A mode with a given frequency will have properties, determined by $\bar\omega$,
that
change over a typical distance of $r\bar\omega/(m\Omega)$.

The general linear equation governing such disturbances is
$$\hfil\partial_z^2\chi+\partial_z\left[\ln\left({S^2\rho\over\bar
\omega^2-N^2}\right)\right]\partial_z\chi+(\bar\omega^2-N^2)
\Bigl({\chi\over c_s^2}-{\partial_r^2\chi\over\Omega^2-\bar\omega^2}+
\ldots\hfil\eqno(1a)$$
$$\hfil\ldots+\left({m\over
r}\right)^2\chi{7\Omega^2-\bar\omega^2\over(\Omega^2-\bar\omega^2)^2}+
{3\Omega\bar\omega{m\over r}\partial_r\chi\over(\Omega^2-\bar\omega^2)^2}\Bigr)
=0,\hfil$$
where
$$\hfil\chi\equiv {\delta P\over P^{1/\gamma}}\hfil,\eqno(1b)$$
$$\hfil S\equiv {P^{1/\gamma}\over\rho}\hfil,\eqno(1c)$$
and
$$\hfil N^2\equiv z\Omega^2\partial_z\ln S\hfil.\eqno(1d)$$
The use of $\chi$ has no particular significance except that it simplifies
the relevant equations.  $N$ is the Brunt-V\"ais\"al\"a frequency.  It vanishes
at the midplane
of the disk and will be large (compared to $\Omega$) and positive in the disk
atmosphere.
In between it may have any value and will be imaginary in convectively unstable
regions.
The associated velocity perturbations are
$$\hfil v_r={i\bar\omega S\over\bar\omega^2-\Omega^2}\left(2\Omega{m\over
r\bar\omega}\chi+\partial_r\chi\right),\hfil\eqno(2a)$$
$$\hfil v_\theta={-S\over\bar\omega^2-\Omega^2}\left(\bar\omega{m\over
r}\chi+{\Omega\over2}\partial_r\chi\right),\hfil\eqno(2b)$$
$$\hfil v_z={i\bar\omega S\over\bar\omega^2-N^2}\partial_z\chi\hfil\eqno(2c)$$
and the associated density perturbations are
$$\hfil
{\delta\rho\over\rho}={\partial_zS\over\bar\omega^2-N^2}\partial_z\chi.\hfil
\eqno(2d)$$
A qualitative sense of the solutions for these equations may be obtained by
replacing the first two terms
in eq. (1a) with $-k_z^2\chi$.

Equations (1) and (2) describe linear perturbations, but in order to estimate
the consequences
of various linear processes we need to have some way of estimating the
nonlinear saturation
of these modes.  In general this can be quite difficult, and requires a
detailed examination
of the mode-mode couplings.  In the particular case of rapid, small scale
strong turbulence we will
adopt the usual heuristic approach of assuming that the effect on other modes
can be modeled
by a turbulent diffusion tensor of the form $D_{ij}=\langle
v_iv_j\tau_{corr}\rangle$.  If
the turbulence is isotropic then we can approximate this by $D\delta_{ij}$
where $D$ is the
mean square velocity of the turbulence times the eddy turnover time.  There is
an additional
dissipative term which is unique to shearing environments (Vishniac and Diamond
1989).  Radial
diffusion coupled to differential rotation will smear wave motions over an
azimuthal
wavelength at a rate of
$$\hfil \tau_{SD}^{-1}\sim\left(\left({m\over r}\right)^2\Omega^2
D_{rr}\right)^{1/3}.\hfil\eqno(3)$$
We will refer to this process as the shear decorrelation rate.  It can be
important
in setting limits on the importance of high $m$ modes in accretion disks.

The angular momentum flux, ${\cal L}_z$ is given by $\rho r\langle
v_rv_\theta\rangle$.  On physical
grounds we expect that for axisymmetric waves this is entirely due to mixing.
Since turbulent
mixing of a conserved quantity tends to minimize its gradient, and since
specific angular momentum
increases outward in a thin accretion disk, the net effect will be to move
angular momentum
{\it inward}.  In fact, from eqs. (2a) and (2b) we see that for $m=0$
$$\hfil\langle v_rv_\theta\rangle ={-S^2\Omega\over
2|\bar\omega^2-\Omega^2|^2}|\partial_r\chi|^2 \Re (i\bar\omega),\hfil\eqno(4)$$
{\elevenit i.e.} growing modes have ${\cal L}_z<0$ as expected.

The energy flux associated with disturbances in the
disk can be shown (cf. Vishniac and Diamond 1989) to have the form
$$\hfil F=\langle \delta P v_r\rangle +\rho r \Omega\langle v_r
v_\theta\rangle+{\cal E}\langle \rho v_r\rangle,\hfil\eqno(5)$$
where ${\cal E}$ is the energy density.  The last term describes the advection
of energy by mass transport
and is of no particular concern.  The first one is the transport of energy
expression that one would
obtain in the absence of a strong zeroth order flow.  It gives the transport of
energy measured by a local
observer, {\elevenit i.e.} terms that go as the square of the perturbation
variables.
This term points in the
same direction as the wave group velocity, as one would expect since it
represents the
transport of a positive quantity by the waves.  The second term is just
${\cal L}_z\Omega$.  Waves carrying angular momentum will slow down,
or speed up, the fluid while they pass, and this gives rise to an energy
contribution proportional
to the product of $r\Omega$ and $v_\theta$.  Since ${\cal L}_z$ can have either
sign the total energy
flux may, in general, have either sign as well, {\elevenit i.e.} a wave
propagating
inward may have a positive
total energy flux.  To linear order both $F$ and ${\cal L}_z$ are preserved as
a wave propagates.
This has the consequence that $\langle \delta P v_r\rangle$ is not.  It is
particularly
important to note that if $\langle\delta Pv_r\rangle$
and ${\cal L}_z$ have opposite signs then as a wave propagates inward $\Omega$
will increase and
so will $\langle \delta Pv_r\rangle$.  Under most circumstances $\langle
v^2\rangle$ will as well.

It is also interesting to note that the condition that ${\cal L}_z>0$ (or in
other words, that the
waves carry angular momentum outward, thereby driving the evolution of the
accretion disk) is
that $\bar\omega^2$ increase in the direction of wave propagation.  Inward
propagating waves
are then particularly helpful since they have $\langle \delta P v_r\rangle<0$,
${\cal L}_z>0$,
and, since $\partial_r\Omega<0$ and $\bar\omega=\omega+m\Omega$, will tend to
grow in amplitude as they
move inward.  This should promote nonlinear dissipation and the deposition of
negative angular
momentum into the disk fluid.  This is the basis for theories of angular
momentum
transport using compressional modes (Shu 1976) and internal waves (Vishniac and
Diamond 1989).
We will now examine each in detail.
\vskip 0.6cm
\line{\elevenit 2.1. Sound Waves\hfil}
\vskip 0.4cm
When perturbations have large frequencies, in the sense that
$\bar\omega^2\gg \Omega^2$, then neglecting vertical structure we can write
$$\hfil \bar\omega^2\approx c_s^2k^2+\hbox{WKB terms},\hfil\eqno(6)$$
{\elevenit i.e.} the usual one for sound waves.
The WKB terms represent the secular effect of the shearing environment
on the wave amplitude, a point we have already mentioned in connection with
the separate conservation of energy flux and angular momentum flux.  A detailed
examination of these waves shows there is a lower limit to
$\bar\omega^2$ for a given $k_z\ne 0$ and $m$.  As a sound wave reaches this
limit it will undergo reflection.  No corresponding upper limit appears.  It
follows that a sound wave
has, at most, one surface for radial reflection.  It appears from this, and
from the preceding
discussion, that one might construct a model
of angular momentum transport based on these waves in the following manner.  If
there is some source
for sound waves at the outer disk boundary, that produces waves that can travel
through the disk without reflection and that have ${\cal L}_z>0$ then nonlinear
interactions between the
disk and sound waves may allow the deposition of negative angular momentum and
the inward transport
of mass.  This is, in essence, the proposal of Shu (1976) who pointed out that
the accretion stream
creates a shock wave when it hits the disk, and the resulting spiral shock may
penetrate deeply
into the disk.  Of course, the properties of such a shock depend to a large
degree on nonlinear terms neglected
in eq. (1) nevertheless it can still be used as a crude basis for understanding
the results of more
detailed calculations.  Following Shu's original suggestion, analytical work by
Donner (1979), Spruit (1987) and
Larson (1990) has shown that the dimensionless value of $\alpha$ obtained in
this way scales
as $(H/r)^{3/2}$. This is significantly larger than the estimate one would
obtain from examining
the angular momentum flux carried by a sound wave and reflects the ability of
coherent, nonlinear effects
to enhance the efficiency of angular momentum transport.
A number of numerical studies (Sawada, Matsuda, and Hachisu, 1986; Sawada,
Matsuda, Inoue and Hachisu, 1987;
Spruit, Matsuda, Inoue and  Sawada 1987; Rozyczka and Spruit 1989; Matsuda,
Sekino, Shima, Sawada, and Spruit 1990)
have confirmed that spiral shocks
result from both tidal effects and the impact stream, and that the shock waves
penetrate to very
small radii.

In spite of the many attractive features of this model, several problems
remain.
First, almost all this work has been done in two dimensions.  The only
exception has been
Sawada and Matsuda (1992) who ran a 3 dimensional simulation of the formation
of a disk
from an accretion stream far enough to see the formation of the spiral shocks,
but not nearly
far enough to show a steady state solution.  In a two dimensional disk $N^2=0$
and $c_s$ is
a constant.  We see from eq. (1) that the resulting linear solutions are almost
trivial and
neglect several physically important features.  Moreover they have the
unpleasant feature that
since $\chi$ is not a function of $z$ the velocity perturbations $\propto
c_s\chi P^{(1/\gamma)-1}$
diverge in an isothermal atmosphere.   This is the result of refraction, a
point examined by Lin, Papaloizou
and Savonije (1990).  They showed that generic compressional disturbances
inside an accretion
disk will tend to quickly reach the disk atmosphere and undergo rapid nonlinear
dissipation.
Realistic solutions to eq. (1) will also show this effect, but have a
compensating feature.
As mentioned above, all waves undergo reflection as $N^2$ increases past
$\bar\omega$.  This
means that a global sound wave with an azimuthal wave number $m$ will undergo
reflection at
$\sim m$ scale heights.  For $m$ small this will keep the wave from dissipating
in the disk
atmosphere.  Unfortunately, this advantage will be lost as the wave steepens
into a shock
and its characteristic time scale falls to zero.  We can only conclude that
vertical confinement
of the wave energy seems to be a serious problem.

Second, this process is relatively ineffective.  The scaling of $\alpha$ with
$(H/r)^{3/2}$ is consistent
with most observational constraints, but the coefficient in front is quite
small, $\sim 0.01$ (Spruit 1987).
The result is a level of angular momentum transport which cannot be easily
reconciled with our
current understanding of dwarf novae.  Some numerical work suggests that this
low value is
not correct (Matsuda, Sekino, Shima, Sawada, and Spruit 1990), but more recent
work by Rozyczka
and Spruit (1992) is consistent with the earlier analytic estimates.

Third, Livio and Spruit (1991) have examined the outburst intervals of dwarf
novae
as a function of mass ratio as a test of the scaling of $\alpha$ with $(H/r)$
and concluded that a gentler
dependence might be preferred.  Evidently there are few grounds for believing
that spiral
shocks mediate angular momentum transport in dwarf novae systems.  If we
conclude that this
process does not dominate in dwarf novae then it becomes hard to understand why
it should dominate in other, less well
observed, systems.
\vskip 0.6cm
\line{\elevenit 2.2. Internal Waves\hfil}
\vskip 0.4cm
The second set of hydrodynamic modes are internal waves, analogous to deep
ocean waves.
For small $m$ they have a local dispersion relation
$$\hfil\bar\omega^2=\left({N^2k_r^2+\Omega^2k_z^2\over
k^2}\right).\hfil\eqno(7)$$
This is identical to the dispersion relation for deep ocean waves, except that
we have replaced the coriolis parameter with $\Omega^2$.  In practice, the
difference
is that in the ocean
we normally consider the coriolis parameter to be small, whereas $\Omega^2$ is
a large
parameter for internal waves in an accretion disk.
We note that in the limit $N^2=k_z=0$ this dispersion relation collapses.
Purely two dimensional treatments of disk structure will not include these
modes.
For three dimensional disks this dispersion relation implies that internal
waves are confined
between a radius of reflection, where $\bar\omega^2\rightarrow\Omega^2$ and
$k_r^2\rightarrow 0$
and a radius of resonant absorption where $\bar\omega^2\rightarrow0$ and
$k_r^2\rightarrow\infty$.
Everywhere in between internal waves are confined vertically to regions where
$\bar\omega^2>N^2$.
As a wave approaches reflection the group velocity implied by eq. (6) is
proportional to the
square root of the distance to the reflection point.  Consequently reflection
takes a finite amount
of time, as expected.  Near reflection, a wave occupies a broad slice of the
disk midplane, although
it still cannot propagate into regions where $N^2>\Omega^2$.  At the other
extreme, as a wave
approaches resonance its group velocity is proportional to the square of the
distance to the
resonant radius.  Consequently the wave energy piles up and undergoes nonlinear
dissipation.  This is exacerbated by the fact that as a wave approaches
resonance it is confined to an increasingly
thin zone around the midplane.
We can conclude from this that in general these waves are transient events that
are confined
to annuli with a width of order $r/m$.  Regardless of starting conditions they
eventually
enter a state of rapid nonlinear dissipation, close to their corotation radius.
 Stationary modes
based on eqs. (1) or (6) are unphysical in that they contain an infinite amount
of energy at
their corotation radius.

This generic picture does {\elevenit not} apply to modes with $m=0,\pm 1$.  The
former
case is of limited interest, since such modes contain no angular momentum (and
are also
incapable of driving a dynamo).  In the latter case, it is possible to arrange
the parameters
of the waves so that their resonant points lie outside the disk and their
reflection point
is at $r=0$.  To be more specific, if we choose waves with $\omega$ less than
$\Omega$ at
the outer edge of the disk and require $m$ and $\omega$ to have opposite signs,
then
$\bar\omega=\omega-\Omega$ fulfills these conditions.  Such waves will be
global modes
of the disk.  This allows us to return to the picture we discussed earlier, in
which angular momentum is transported
outward by inward propagating waves whose linear amplification leads to strong
nonlinear interactions and the deposition of negative angular momentum at small
radii
(cf. Vishniac and Diamond, 1989).
It turns out that in conducting disks this effect is overwhelmed by the role of
these
waves in driving a dynamo.  Still it is worth discussing some of the details of
the
propagation of these waves, both to compare the results to the spiral shock
wave model,
and to set the stage for a discussion of the wave-driven dynamo model.

Assuming, for the moment, such waves are excited at large amplitude at the
outer edge of the disk, what
happens to them as they move inward?
First, they will undergo
linear amplification at a rate of $V_{group}/r$.  Modes with small $m$ and
$\bar\omega/\Omega$ of order unity have
$V_{group}\sim c_s(k_zH)^{-1}$.
The amplification will also be driven by geometric
focusing as $H$ and $r$ decrease together, although this will increase the
result only by a factory of order unity.  In a steady state this increase will
be balanced by
nonlinear dissipation.  The nonlinear dissipation of such waves is complicated.
 The radial
group velocity has a strong dependence on $\bar\omega$ so we expect that
incoherent nonlinear interactions
will be important.  This suggests a generic nonlinear dissipation rate of
$$\hfil\tau_{decorr}^{-1}\sim {k_z^2\langle v_z^2\rangle +k_r^2\langle
v_r^2\rangle\over\bar\omega}\hfil\eqno(8)$$
which is consistent with work on deep ocean waves (McComas and
Bretherton 1977; M\"uller {\elevenit et al.} 1986)
There may be important coherent effects as well, {\elevenit e.g.} the
formation of solitons, but little is known
about this possibility or its consequences.  Equating the linear amplification
rate with $\tau_{decorr}^{-1}$
then for $\bar\omega$ some fraction of $\Omega$ we find that
$$\hfil\langle v^2\rangle\sim {H\over r} c_s^2(k_zH)^{-3}.\hfil\eqno(9)$$
We see that the wave spectrum will be dominated by modes with $k_zH\sim 1$.
The corresponding angular momentum flux is
$$\hfil {\cal L}_z\sim {H\over r}c_s^3.\hfil\eqno(10)$$
Assuming that nonlinear dissipation deposits the angular momentum in the waves
into the disk
at a rate of ${\cal L}_z \tau_{decorr}^{-1}/V_{group}$ gives an `alpha'
equivalent to
$$\hfil\alpha\sim (H/r)^2\hfil\eqno(11)$$

There is a major assumption contained in these estimates, that the nonlinear
interactions between the
various modes lead to a steady state with $\langle
\bar\omega^2/\Omega^2\rangle$ a constant less
than one everywhere inside the disk.  If this fails then
$\bar\omega\rightarrow\Omega$ as the waves
go to small radii and the group velocity and amplitude will both decline
sharply.  Simulations of
the evolution of the internal wave spectrum based on the weakly resonant
approximation and assuming
random phases supports this assumption (Huang 1992), but more realistic
calculations are still desirable.

Is it fair to assume that internal waves of this strength are excited at the
outer edge of the disk?
It is true that waves of smaller strength will grow to saturation given enough
time,
{\elevenit i.e.} a large
enough number of radial e-foldings, but this growth goes as $r$ to a modest
power and will be insufficient
if the waves are very weak at the outer edge of the disk.  At this point the
excitation of internal waves is not understood,  and there are a
number of unresolved issues associated with this problem.  A
non-self-gravitating
disk in a binary system has two possible sources of negative angular momentum,
tidal interactions
and the accretion stream.  The former is generally supposed to be the ultimate
sink of excess
angular momentum from the disk, and will involve a loss of angular momentum at
a rate of
$\sim \dot Mr^2\Omega$.  The latter is due to the addition of material with a
specific angular
momentum far below the average value at the outer edge of the disk, so that the
stream will
constitute a local sink of angular momentum with a total rate comparable to the
tidal loss
rate.  An ingoing stream of $|m|=1$ internal waves with an amplitude of
$\langle v^2\rangle\sim (H/r)c_s^2$
(see below) will carry an angular momentum flux of approximately $\Sigma
H^2c_s^2$.  This
implies that either the accretion stream or tidal interactions need to excite
ingoing $|m|=1$
internal waves with an efficiency of $\sim (r/H)^2\alpha$.  Goodman  (1992) has
suggested
that a standing wave pattern of $m=-1$ waves should result from the action of
the $m=2$ component
of the tidal forces, but this may be sensitive to the inner and outer boundary
conditions of the
disk.  Lubow (1992) has pointed out that tidal forces will also excite vertical
motion within the
disk, which corresponds to the excitation of the minimal $k_z$ mode of internal
waves, but
no quantitative results are available.
In two dimensional models the impact stream excites modes with
$\omega=\omega_{binary}$ and
$k_z=0$ ({\elevenit i.e.} $\chi$ not a function of $z$), but in real life a
substantial component
of the forcing will have $\chi$ a function of $z$ and drive internal waves.
This question
has not been properly addressed since previous treatments of the interaction
of the disk with the accretion stream and/or the tidal forces from the
secondary star have been
almost entirely two dimensional.  Even granting the existence of one, or more,
effective excitation
mechanisms there is also the issue of how quickly high amplitude, coherent
waves excited near
the outer edge of the disk evolve into the expected low amplitude, incoherent
waves we expect at
smaller radii.  In sum, there seems to be plausible grounds to suspect that
these waves are excited,
but almost nothing is actually known about the process.  We note that the issue
here goes well beyond the
existence of the waves.  If tidal forces are ineffective at exciting them, but
the accretion
stream provides a strong excitation mechanism, then temporary interruptions in
the accretion stream
ought to result in a rapid cooling of the disk.  Such interruptions seem to
occur frequently in
cataclysmic variables just above the period gap (Robinson, Barker, Cochran,
Cochran and Nather 1981),
and do result in a rapid descent into an extremely quiescent state.  If we
consider an isolated disk then
the situation is even murkier.  Adams, Ruden, and Shu (1989) and Shu, Tremaine,
Adams, and Ruden (1990)
have shown that massive disks ({\elevenit i.e.} disks with a total mass
comparable to the central object) have a fairly
robust $m=1$ instability centered on a narrow annulus at
the outer edge of the disk.  More recently Noh, Vishniac, and Cochran (1992)
showed that a variant of this instability
can persist down to very low mass disks, for which $M_{disk}\sim (H/r)M_*$,
{\elevenit i.e.}
disks for which our
(defining) constraints on mass are marginally violated. Unfortunately the
coupling between
this instability and ingoing internal waves has not been explored. In any case,
isolated disks for which
$M_{disk}\ll (H/r) M_*$ even at the outer edge would seem to be bereft of any
exciting mechanism.
Fortunately, the notion that such disks are completely inactive seems
consistent with all the available
observational evidence.

How does this estimate compare to the one derived from the theory of spiral
shock waves?
The scaling law presented here is steeper, which may be even more of a problem
for dwarf novae
models (cf. Livio and Spruit).  On the other hand, the coefficient, although
uncertain, ought to
be of order unity, suggesting that this effect will dominate for disks with
$(H/r)$ greater
than about $10^{-4}$.
The vertical structure of the disk poses no particular problem for internal
waves, in contrast
to the difficulties and uncertainties involved with sound waves and shocks.
On the other hand, we have already noted that the excitation of internal waves
is much less well understood.
Finally, the complicated nature of the nonlinear evolution of the internal
waves
simply makes it more difficult to provide reasonable analytical estimates of
the coefficients
of the scaling laws in this case.
\vskip .6cm
\line{\elevenit 2.3. Convection\hfil}
\vskip .4cm
It is immediately apparent from eq. (7) that when $N^2<0$ an instability is
lurking in the dispersion relation for internal waves, {\elevenit i.e.}
standard convection in
a shearing environment.  The presence of shear poses
no particular obstacle to this instability (although it does limit it in some
interesting ways) and so we would expect convective turbulence to result.
This has been invoked many times as the basis for a theory of angular momentum
transport (Cameron, 1969; Paczynski, 1976; Vila, 1978; Lin and Papaloizou,
1980;
Lin, 1981; Smak, 1982; Cabot {\elevenit et al.}, 1987a,b).

Although it is clear that convective mixing ought to transport angular
momentum,
there are some difficulties with using it as the basis for a {\elevenit
general} model
of angular momentum transport.  For example, convection will continue only if
the entropy of the
gas continues to decrease away from the midplane.  The entropy gradient is, in
turn,
a function of opacity and the vertical profile of heat deposition.  In this
model, heating is due to
the dissipation of orbital energy, which is, in turn, due to convective mixing.
 There is
no obvious reason why this should form a self-consistent cycle.  Also,
optically thin,
nearly isothermal disks seem to exist, and in such disks a convective cycle
will not
start, and if started will quickly die away.

Perhaps the central question concerning convection, however, is that of the
{\elevenit direction}
of angular momentum transport. For example, Goodman and Ryu (1992) have
asserted
that convection invariably transports angular momentum inward in accretion
disks, on the basis of
a shearing sheet calculation in which $k_r\rightarrow\infty$ as
$t\rightarrow\infty$.
This is clearly a crucial issue, as only outward transport will
facilitate the accretion process.  We have already noted that for axisymmetric
modes
angular momentum is transported inward in an accretion disk.
The ambiguity in the sign of ${\cal L}_z$ is a natural
consequence of the fact that in an accretion disk, the driving entropy gradient
is vertical,
not radial.  Hence, only vertical entropy mixing follows from the standard lore
concerning
the relation between gradients and the instabilities they cause.  The intrinsic
difficulty in determining the angular momentum flux, as well as its relation to
the convective
cell dynamics, is nicely illustrated by considering Rayleigh-Bernard convection
in a thin
disk in solid body rotation, with $|N|^2$ of order $\Omega^2$.  In this case
the convective
growth rate is
$$\hfil \Gamma^2={(k_r^2+(m/r)^2)|N|^2-k_z^24\Omega^2\over
k_r^2+k_z^2+(m/r)^2},\hfil\eqno(12)$$
where we have replaced $\bar\omega^2$ with $-\Gamma^2$ since we are dealing
with an instability
whose frequency is no longer a function of $r$.  The quantity
$4\Omega^2=2\Omega r^{-1}\partial_r(r^2\Omega)$
is the Rayleigh discriminant (equal to $\Omega^2$ for a Keplerian disk).
Equation (12) expresses the
well-known relation between convective cell structure and instability enforced
by the Taylor-Proudman
theorem for a rapidly rotating fluid.  Instability requires that the vertical
cell scale associated
with vertical vortex line bending not be less than the planar cell scale,
{\elevenit i.e.} $k_z^2<k_r^2$ or $(m/r)^2$.
Note that cells with $k_r^2>k_z^2\gg (m/r)^2$ are shaped like toroidal rolls
with an
ellipsoidal cross section in the $r-z$ plane, while cells with
$(m/r)^2>k_r^2\gg k_z^2$
resemble anisotropic Proudman Pillars, with an elongated ellipsoidal cross
section
lying in the $r-\theta$ plane.  The impact of this eddy anisotropy on angular
momentum
transport is in turn illustrated by the quasi-linear expression for the angular
momentum flux, which
is:
$$\hfil{\cal L}_z=\rho
r{S^2\over(\Gamma^2+4\Omega^2)^2}\left(2\Omega\Gamma\left(\left({m\over
r}\right)^2-k_r^2\right)-(4\Omega^2-\Gamma^2){m\over
r}k_r\right)|\chi|^2\hfil\eqno(13)$$
The first two terms are primarily due to coriolis force effects, while the
third is related to
pressure gradient induced transport.  The
resulting angular momentum transport is clearly very sensitive to eddy
anisotropy, in that
for $k_r\ll (m/r)$ (corresponding $v_r\gg v_\theta$ for incompressible flow
with $k_z$ small)
transport is outward, while for $(m/r)\ll k_r$ (corresponding to $v_r\ll
v_\theta$)
transport is
inward.  Isotropic convection cells make little or no contribution to ${\cal
L}_z$.
We note that the third term is, in general, quite significant, although unless
the symmetry between $r$ and $\theta$ is broken, contributions from modes with
opposite helicity will cancel. We see that the sign of ${\cal L}_z$ is
determined by the
detailed structure of the eddy wave-number spectrum, which is difficult to
predict.

In a thin disk differential rotation distorts finite $k_z$ modes into locally
unstable spiral modes.
We expect that the linear eigenfunctions of eq. (1) for this case will be
similar to the non-rotating case sufficiently close to the corotation point
that
$|\Re\bar\omega|\ll\Gamma$, {\elevenit i.e.} with a scale length of
$\sim \Gamma r/(m\Omega)$.
At larger distances the instability will look like outgoing internal waves and
the
amplitude will fall off rapidly.  The net angular momentum flux is
$$\hfil{\cal L}_z=\rho
r\langle{S^2\over|\bar\omega^2-\Omega^2|^2}\left(\Omega\Gamma\left(2
\left({m\over r}\right)^2|\chi|^2-{1\over 2}|\partial_r\chi|^2\right)+(\Omega^2
-|\bar\omega|^2)i{m\over r}\chi^*\partial_r\chi\right)\rangle.\hfil\eqno(14)$$
We note that the symmetry between $(m/r)$ and $k_r$ is explicitly broken in
this expression.
Moreover, since $r$ and $\theta$ are no longer symmetric, there is no reason to
expect
that the third term will tend to sum to zero.  Given that the peak of
$|\chi|^2$ is
confined within an annulus of size $\Gamma r/m\Omega$ we can see that the
second term
is always negligible.  In this sense, convection in accretion disks does seem
to favor
inward transport of angular momentum.  Unfortunately, when $m/r$ is large, so
that the radial
wavelength of the instability fills the convective annulus, then the third term
will
will be comparable to the first.  Since $k_r$ does not become arbitrarily large
as $t\rightarrow\infty$
in the normal mode analysis, the exact sign of ${\cal L}_z$ for a given mode
can only be determined by explicit calculation.  Moreover, since large $m$
modes will
be favored by phase space arguments, and not more subject to dissipation unless
$m\gg rk_z$,
the general direction of angular momentum transport due to convection in
accretion disks
must be regarded as uncertain.  It is interesting to note that this question is
related
to the question of whether or not differential rotation inhibits convective
instabilities.
On energetic grounds we expect that if ${\cal L}_z<0$, so that differential
rotation acts
as a sink of energy in convective regions, then convective instabilities will
be
inhibited in Keplerian disks.

Finally, convection will not
be terribly effective as a source for $\alpha$ unless it is very strong.  We
note from
eq. (7) that coriolis forces limit the radial wavelength of a convective cell
to less than $\sim k_z^{-1} \Gamma/\Omega$, where $\Gamma\equiv iN$.  For weak
convection
the convective cells will be tall and thin.  Approximate incompressibility
implies
that $v_r\sim (k_z/k_r)v_z$ and on dimensional grounds we expect $v_z\sim
v_\theta\sim k_z^{-1}\Gamma$.
These estimates are self-consistent in the sense that using them we obtain
$D_z\sim k_z^{-2}\Gamma$
and $D_r\sim k_r^{-2}\Gamma$ so that the turbulent damping rate is comparable
to $\Gamma$.
These estimates also assume that $(m/r)$ is at least not very much larger than
$k_z$.
Similar arguments suggest that very large $m$ modes will tend to be relatively
ineffective in driving transport.  On the other hand, it is well to remember
that
this discussion is implicitly quasi-linear, and real convective turbulence
shows
features that are not easily understood in such simple terms.
As a best estimate we conclude that
$\alpha\sim \langle v_rv_\theta\rangle/c_s^2\sim (\Gamma/\Omega)^3$
in agreement with Ruden, Papaloizou and Lin (1988) (although their discussion
was based on axisymmetric modes for which $\alpha$ is necessarily negative).
We stress again that the actual sign of $\alpha$ in this case is uncertain.

We note that Korycansky (1992) has suggested that an infinitesimal amount of
dissipation
will eventually suppress any convective mode, albeit after an extended period
of growth.
This is an artifact of relying on a plane wave approximation for the
radial structure of the convective modes, which includes the unfortunate
consequence
that $k_r\propto t$ and the dissipative damping rate $k_r^2D\rightarrow\infty$
as $t\rightarrow\infty$.

In view of these problems we tend toward the view that while convection may
have some
effect on disk structure, it is probably limited to a subsidiary role.  A
general
model for $\alpha$ must come from some other mechanism.
\vskip .6cm
\line{\elevenit 2.4. The Magnetic Shearing Instability\hfil}
\vskip .4cm
Inasmuch as ionized disks are extremely well conducting systems
it seems reasonable to explore the possibility that they contain some
magnetic field.  Such a magnetic field has been suggested as the source
of angular momentum transport in disks by a number of authors ({\elevenit e.g.}
Lynden-Bell
1969; Shakura and Sunyaev 1973; Eardley and Lightman 1975; Ichimaru 1977;
Coroniti 1981).
Due to the presence of differential rotation
any radial field component  will lead directly to a rapid increase
in $B_\theta$.  It follows that we should expect the magnetic field
to be predominantly azimuthal.  Perturbations of a highly conducting disk with
an azimuthal field will have a fairly complicated dispersion relation.
Rather than present the generalization of eq. (1), which is not
particularly enlightening in any case, we will take a simplified case.
If we assume incompressibility and $N^2=0$ the dispersion relation
becomes (cf. Vishniac and Diamond, 1992)
$$\hfil\left(1-{k_z^2\over
k_*^2}{\Omega^2\over\tilde\omega^2}\left(1+4{\omega_A^2\over\tilde\omega^2}
\right)+{9\over2}{(m/r)^2\omega_A^2\Omega^2\over k_*^2(\tilde\omega^2+
\omega_A^2)\tilde\omega^2}\right)v_r-{1\over k_*^2}\partial_r^2v_r=
-3{\omega_A^2\over\tilde\omega^2}{\Omega\over\bar\omega}{m\over rk_*^2}
\partial_rv_r\hfil\eqno(15a)$$
where
$$\hfil\tilde\omega^2\equiv\bar\omega^2-\omega_A^2\hfil\eqno(15b)$$
$$\hfil\omega_A\equiv V_A{m\over
r}\equiv\left({B_\theta^2\over4\pi\rho}\right)^{1/2}{m\over r}\hfil\eqno(15c)$$
and
$$\hfil k_*^2\equiv k_z^2+\left({m\over r}\right)^2.\hfil\eqno(15d)$$
The other hydrodynamic perturbation variables are
$$\hfil v_\theta={ik_z^2\Omega\over
k_*^22\bar\omega}\left(\tilde\omega^2+4\omega_A^2\over\tilde\omega^2\right)v_r
+i{m\over rk_*^2}\partial_rv_r\hfil\eqno(16a)$$
$$\hfil v_z={-ik_zm\Omega\over
k_*^2r2\bar\omega}\left(\tilde\omega^2+4\omega_A^2\over\tilde\omega^2\right)v_r
+i{k_z\over k_*^2}\partial_rv_r\hfil\eqno(16b)$$
and
$$\hfil\delta P=\rho\left({i\Omega r\over 2mk_*^2}v_r\left(\left({m\over
r}\right)^2-k_z^2{4\omega_A^2\over\tilde\omega^2}\right)-i{\bar\omega\over
k_*^2}\partial_r v_r\right),\hfil\eqno(16c)$$
where $V_A\equiv (B^2/4\pi\rho)^{1/2}$ is the Alfv\'en speed.
The magnetic field perturbations have the form
$$\hfil \vec b=B_\theta{m\over r\bar\omega}\vec v+iB_\theta{3\Omega m\over
2\bar\omega^2 r}v_r\hat\theta.\hfil\eqno(16d)$$
A somewhat more straightforward dispersion relation results if we assume that
we
are in the limit of large radial wavenumbers, {\elevenit i.e.}
$k_r\gg m\Omega/(r\bar\omega)$ so that we can approximate the radial
eigenfunctions as plane waves.  Then the RHS of eq. 12a can be neglected
and the surviving terms can be written in the form
$$\hfil\tilde\omega^6+\tilde\omega^4(\omega_A^2-\omega_I^2)+\tilde\omega^2(
-5\omega_A^2\omega_I^2+{9\over2}{(m/r)^2\over k^2}\omega_A^2\Omega^2)
-4\omega_I^2\omega_A^4=0,\hfil\eqno(17a)$$
where
$$\hfil\omega_I^2\equiv {k_z^2\over k^2}\Omega^2.\hfil\eqno(17b)$$
In the limit where $\omega_A^2\ll\omega_I^2$ and $m/r\ll k_z$ eq. (17) has a
root with
$\bar\omega^2\approx 3(2(m/rk_z)^2-1)\omega_A^2$, {\elevenit i.e.} an
instability.
Some discussion of this case using a radial plane wave WKB analysis can be
found
in Balbus and Hawley (1992).

This instability was discovered by Velikhov (1959) and
described at length by Chandrasekhar (1961) in the context of a vertical
magnetic
field threading an incompressible Couette flow.  We will refer to it here as
the
Magnetic Shearing Instability (MSI).  Its novel feature is that it depends
only on $\partial_r\Omega<0$, rather than an unstable gradient
in the specific angular momentum, and necessarily acts to transport angular
momentum in the direction
of decreasing {\elevenit angular speed}.  It is also present in the limit of an
arbitrarily weak
magnetic field, although its transport effects become negligible as $\vec
B\rightarrow 0$.
These novel features follow from the way that the distorted magnetic field
lines serve to
tie together fluid elements at different radii.  The more slowly moving fluid
at larger radii
is accelerated by fluid elements lying on the same field line but at smaller
radii, thereby
transferring angular momentum outward.  In spite of the relatively
early date of this work its importance for the dynamics of accretion disks was
not
realized until recently (Balbus and Hawley 1991; Hawley and Balbus 1991).
Subsequent
work on the nonlinear evolution of this instability (Hawley and Balbus 1992a;
Zhang, Diamond, and
Vishniac 1992) has been largely confined to a consideration of the unstable
axisymmetric modes
of a vertical field.  This turns out to be of limited use in considering the
role of
the MSI in a realistic accretion disk.

What are the critical wavelengths for MSI?  We note from eq. (16) that the
instability
exists for $\omega_A^2<\omega_I^2$, in which case the growth rate is of order
$\omega_A$,
but is suppressed for $\omega_A^2>\omega_I^2$.  This implies that it has a
maximum growth
rate of order $\Omega$, which is achieved only when $k_z$ is not much less than
$|k|$ and
$(m/r)V_A\sim \Omega$.  Apparently the fastest growing modes have azimuthal
wavelengths
of $V_A/\Omega$ and radial and vertical wavelengths which are comparable to
each other,
and no larger than the azimuthal wavelength.  More slowly growing modes can
have much
larger azimuthal and vertical wavelengths.  We note that they cannot
have larger radial wavelengths.  The validity of the plane wave approximation
rests on the
condition that the radial wavelength be less than the characteristic distance
for changes
in $\bar\omega$.  For MSI acting on an azimuthal magnetic field this implies
$$\hfil k_r\gg {\partial_r\bar\omega\over\bar\omega}\sim {(m/r)\Omega\over
\omega_A}\sim {\Omega\over V_A}.\hfil\eqno(18)$$
In other words, MSI modes for an azimuthal field have a (narrow) characteristic
radial scale.
This seems at first to be counterintuitive.  After all, one could always define
initial
conditions with a large radial scale.  Equation (18) follows from the fact that
after
one characteristic growth time differential rotation will have automatically
reduced that
initial scale, no matter how large, to $\sim V_A/\Omega$.  This argument does
not
apply to the MSI modes of a vertical field, which have a growth rate
proportional
to $k_zV_A$ instead.  In that case, the vertical wavelength is limited only by
the thickness
of the disk, and the radial wavelength is limited to a similar value by
secondary
instabilities (Zhang, Diamond, and Vishniac 1992).

What is the angular momentum transport associated with the MSI?
We have
$$\hfil{\cal L}_z=\rho r\langle v_rv_\theta\rangle -r{\langle
b_rb_\theta\rangle\over 4\pi}.\hfil\eqno(19)$$
Using eqs. (15) and (16) we can show that
$$\hfil {{\cal L}_z\over r\rho}=\langle
iv_r^*\partial_rv_r\left(\left(1-{\omega_A^2\over|\bar\omega|^2}\right){m\over
4rk_*^2}\left(1+{3\omega_A^2\over\bar\omega^2}\right)-{3\over2}
{\Re\bar\omega\over\bar\omega}{m\tilde\omega^2\omega_A^2\over rk_*^2
|\bar\omega|^4}\right)+\hfil\eqno(20)$$
$$\hfil\left(1-{\omega_A^2\over|\bar\omega|^2}\right){1\over 2\Omega
k_*^2}{i\tilde\omega^2|\partial_r v_r|^2\over\bar\omega}+\ldots\hfil$$
$$\hfil\ldots+{i\Omega\over
2\bar\omega}|v_r|^2\left(-3{\omega_A^2\over|\bar\omega|^2}+\left(1-
{\omega_A^2\over|\bar\omega|^2}\right)\left({\tilde\omega^2\over\Omega^2}+
{9(m/r)^2\omega_A^2\over2 k_*^2\bar\omega^2}\right)\right)\rangle.\hfil$$
We note that an unstable mode will be localized around the corotation point,
where $\bar\omega$ is purely imaginary,
and will have an amplitude that drops sharply to zero as the real part of
$\bar\omega$ becomes important.  In
this region all the terms comprising ${\cal L}_z$ are positive for a growing
mode, except for the first, whose
sign is ambiguous.  Fortunately, the lower limit on $k_r$ imposed by
differential rotation implies that this
term is smaller than the others.  The magnetic shearing instability transports
angular momentum outward.

Although a detailed evaluation of the nonlinear effects of the MSI has not been
completed for the physically interesting case of an azimuthal magnetic field,
some
qualitative comments are in order.  We can begin by considering the
modes for which $(m/r)\sim k_z\sim k_r\sim \Omega/V_A$ and which consequently
grow at rate $\sim \Omega$. The instability will tend to induce
velocity perturbations of order $V_A$.  This speed is, of course, what one
would expect from
a process driven by magnetic forces, but we also note that this is
sufficient to alter the local value of the shearing thereby eliminating the
driving
force behind the instability.  Since the wavenumbers are all comparable the
local
turbulence induced on these small scales will be approximately isotropic.  The
turbulent diffusion coefficients will all be of order $V_A^2/\Omega$.
Similarly, we can
see from eq. (19) that the angular momentum flux will be $\sim V_A^2$ for
an effective `$\alpha$' of $(V_A/c_s)^2$.  There remains the question of
whether or
not the fastest growing modes are actually the ones that dominate the transport
terms.  More slowly growing modes with longer characteristic scales would seem
to have an advantage.  In the case of an azimuthal field this does not appear
to
be the case.  Since all modes have comparable radial wavelengths, nonlinear
interactions between modes appear in the force equations with an associated
time scale of $k_r^2 D_r\sim (\Omega/V_A)^2 V_A^2/\Omega\sim\Omega$.  For all
but the fastest growing modes this term will totally dominate the dynamics of
motion, destroying the long azimuthal wavelength modes through turbulent
diffusion.
We are led to a picture in which a large scale magnetic field coexists with a
turbulent,
disordered field of comparable amplitude  on scales of $V_A/\Omega$ whose
features change at a rate $\sim\Omega$.

This argument does not apply to the MSI of a vertical field, in which
the slowest modes have $k_r\sim k_z\sim 1/H$ and a turbulent diffusion rate of
$H^{-2}V_A^2/\Omega\sim (V_A/H) (V_A/c_s)$.  In this case, the turbulent
diffusion rate
is smaller than the linear growth rate by a factor of $V_A/c_s$ and the slow
linear
modes are unaffected by small scale turbulence.
Eventually the azimuthal field will build up to the point where its MSI
can suppress the vertical field modes.  This implies a saturation field
strength such that $V_A^2\sim V_{Az}c_s$, with typical turbulent eddy scales
of $\sim (V_{Az}H/\Omega)^{1/2}$. By contrast, in two dimensional numerical
models the
azimuthal field has no effect on the instability and grows until it reaches
$V_A\sim c_s$
(Hawley and Balbus 1991; Zhang, Diamond and Vishniac 1992).
In either case, the resultant angular momentum transport
is equivalent to an $\alpha$ of $\sim V_{Az}/c_s$ (Zhang, Diamond and Vishniac
1992).
Preliminary results from three
dimensional simulations are consistent with our prediction of a low saturation
value for $V_A$ (Hawley and Balbus 1992b), but have insufficient dynamic range
to provide a conclusive test.
(We note in passing that these authors take an opposing view of the
interpretation
of their numerical experiments.)

The MSI seems exceptionally well suited to angular momentum transport in an
accretion disk.  It
is a robust instability which moves angular momentum in the right direction.
It even allows us
to estimate the dimensionless viscosity.  The work presented above, however, is
not a
complete predictive theory.  Unless we have a way of understanding the
equilibrium strength of the magnetic field in the disk we have merely
substituted one free parameter,
$\alpha$, for another, $B_\theta^2/P$.  To go further, we need
to understand the accretion disk dynamo.
\vskip .6cm
\line{\elevenit 2.5. Magnetic Buoyancy\hfil}
\vskip .4cm
One important effect associated with vertical motions in a magnetized fluid is
magnetic buoyancy, or the Parker instability.  Although quite general, the
details
of this effect depend on the detailed vertical profile of the fluid and its
derivation
is rather involved.  Here we will simply quote some relevant results and refer
the
reader to Parker (1979) for a detailed treatment of this process.

The basic physical driving mechanism for this instability is the buoyant force
associated
with the fact that gas with an associated magnetic field can maintain a higher
pressure,
for a given density, then unmagnetized gas.  In a magnetized fluid rising field
lines can
shed mass if other sections of the same field lines are sinking, so that gas
frozen to
the field lines can slide off the rising magnetic bubbles.
Depending on the detailed vertical profile
of the magnetic field and the gas this instability will lead to a bending mode
of the
magnetic field lines.  This instability can actually be enhanced in the
presence
of high thermal conductivity.  The vertical wavelength of the unstable modes is
on the order of a scale height.  The growth rate can be as high as $V_A/H$ for
modes
with a wavelength parallel to the field lines of order $H$.  The instability is
insensitive to the transverse wavelength.  It follows that this instability can
persist in accretion disks if we restrict ourselves to sufficiently short
radial
wavelengths, {\elevenit i.e.} $k_r\gg \Omega/(H\Gamma)\sim \Omega/V_A$.  This
was confirmed
in a detailed calculation by Shu (1974).

The usual picture of this instability is
that it leads to flux expulsion at a rate of $V_A/H$ as rising magnetic loops
reconnect in the disk atmosphere.  Vishniac and Diamond (1992) pointed out that
this may be extremely misleading.  Although magnetic buoyancy persists in the
linear theory of
MHD perturbations in accretion disks, the usual dimensional estimates of
vertical
mixing and magnetic flux ejection depend on the saturation state of this
process.  However, in accretion disks a significant complication arises.
The minimum radial wavenumber associated with magnetic buoyancy is, as noted
above,
$\sim \Omega/V_A$.  This wavenumber is virtually the same as the one associated
with the MSI so we can expect a close interaction between the two sets of
modes.
Worse, the Parker instability is relatively slow for $V_A\ll c_s$.  If we
assume
that the action of the MSI on the Parker instability can be represented by a
kind
of turbulent diffusion, then the associated diffusive term gives a damping rate
of $\sim k_r^2 (V_A^2/\Omega)\sim \Omega$.  Physically, what is happening is
that
the MSI modes are causing
the rising segments of field lines to trade momentum with nearby sinking
segments
on a time scale much shorter than the growth time for the instability.
Of course, the buoyancy of the field cannot be entirely suppressed, but
the associated growth rate is reduced to
$\Gamma\sim(\Gamma_{Parker}^2/(k^2D))\sim(V_A/c_s)^2\Omega$.
Assuming efficient reconnection, this rate is the rate at which the Parker
instability can eject magnetic flux from the disk.  This revised estimate is
not only much smaller than one obtains by ignoring MSI, it is also
comparable to the rate at which turbulent diffusion (due to MSI) will
transport flux.  Evidently the magnetic field buoyancy does not introduce
any new time scales.

We also note that shear decorrelation effects will also suppress the Parker
instability.  From eq. (3) we get for $(m/r)>H^{-1}$
$$\hfil\tau_{SD}>\left(\left({\Omega\over
H}\right)^2{V_A^2\over\Omega}\right)^{1/3}\sim\left({V_A\over
H}\right)^{2/3}\hfil\eqno(c1)$$
This is still faster than the magnetic buoyancy rate, although not as fast as
the radial mixing rate.

This does not amount to a {\elevenit complete} suppression of the Parker
instability in accretion disks.  It will still compete with turbulent diffusion
in determining the vertical gradient of the magnetic field.  Also,
we would still expect the emergence of magnetized bubbles to
contribute to heating the disk atmosphere.  The strong interaction between
the MSI and the Parker instability does imply that
magnetic buoyancy does not act to limit the growth of magnetic fields in disks.
Moreover, the loss of spatial and temporal coherence of the modes of the
Parker instability makes it a poor candidate for the driving mechanism
in a magnetic dynamo, a role that has been previously suggested by
Stella and Rosner (1984), and more recently by Tout and Pringle (1992).
We will return to this point in the next section.
\vskip .6cm
\line{\elevenit 2.6. Boundary Instabilities\hfil}
\vskip .4cm
Although no local hydrodynamic instabilities are evident in eq. (1) it
turns out that one can induce instabilities by attaching the accretion disk
to a rigid radial boundary.  This has been shown by numerous
authors (Drury 1980, 1985; Papaloizou and Pringle 1984, 1985; Goldreich and
Narayan 1985;
Narayan, Goldreich and Goodman 1987; Glatzel 1988) in various ways.
The instability arises from the ability of reflecting
radial boundary conditions to cause over-reflection of the impinging waves
(Goldreich and Narayan 1985),
and vanishes for absorbing boundaries (Blaes 1987).  If
a natural reflection boundary for the waves exists inside the disk, then this
process can be repeated until they shock.  It
is possible that this plays a role in the dynamics of boundary layers in
accretion disks, but it is extremely unlikely that this is important in the
bulk of the disk. Numerical studies ( Glatzel 1986; Goldreich, Goodman and
Narayan 1986;
Hanawa 1988a,b) have shown the effects of this instability are
concentrated near the boundary, while calculations by Kaisig (1989) have
demonstrated that this process saturates in a series of weak shocks that have
only a small effect on angular momentum transport.
\vfill
\eject
\vskip .6cm
\line{\elevenit 2.7. Finite Amplitude Instabilities\hfil}
\vskip .4cm
In spite of the stability of accretion disks to infinitesimal hydrodynamic
disturbances (see also Dubrulle and Knobloch 1992) there remains the
possibility
that some finite amplitude disturbance might manage to perpetuate itself, or
generate
equivalent motions nearby.  This possibility has been explored through work
which
examines the response of the disk to a local defect in the flow (Lerner and
Knobloch 1988;
Dubrulle and Zahn 1991) and also through studies of simple turbulence closure
schemes
(Iroshnikov 1980; Dubrulle and Valdettaro 1992; Dubrulle 1992).  In the first
case one
can see that the defect induces instabilities in the surrounding flow.  The
second
approach has yielded the suggestion that an inverse turbulent cascade may arise
in
accretion disks which prevents the destruction of large scale defects in the
flow
and maintains a moderately strong turbulent flow.  It is difficult to evaluate
the future prospects for this work.  In spite of the work done to date the
persistence of
defects in the flow cannot be rigorously justified.  On the other hand, it is
difficult
to rule out such effects.  Current three dimensional hydrodynamic work has not
confirmed
this model, but problems with dynamic range and artificial viscosity still
limit the
ability of computers to confront these models directly.  If this work is
correct, then
it falls into the category of a local model of angular momentum transport for
which
$\alpha$ has a definite and universal value.
\vskip .6cm
\line{\elevenbf 3. Disk Dynamos\hfil}
\vskip .4cm
The generation of large scale magnetic fields in highly conducting fluids, like
stars or accretion disks, is not particularly well understood, although
it seems clear from stellar magnetic reversals that this process must be
possible.  In what follows we will adopt the formalism of mean-field dynamo
theory, while mentioning unresolved issues associated with this
approach.  Reviews of mean-field dynamo theory can be found in
Moffatt (1978), Parker (1979), and Zel`dovich, Ruzmaikin and Sokoloff (1983).
The basic idea in this approach is that the magnetic field can be divided
into a large scale, slowly evolving component, $\vec B$, and a small scale,
rapidly changing component, $\vec b$.  The local velocity field interacting
with $\vec B$, through the induction equation, will induce some $\vec b$.  This
same velocity field acting on $\vec b$ will produce a change in $\vec B$.  A
large scale, fast dynamo results if there exists a mode with $\vec B$
increasing.
In order to obtain the mean field dynamo equations one iterates through the
induction equation and averages over small distances and times.  We obtain
$$\hfil\partial_t
B_i=\epsilon_{ijk}\partial_j(\alpha_{kl}B_l)+\partial_i(D_{ij}\partial_j
B_i)-\partial_j(D_{ik}\partial_k B_j),\hfil\eqno(21a)$$
where $\epsilon_{ijk}$ is antisymmetric under the interchange of indices,
repeated indices implies summation and
$$\hfil\alpha_{kl}\equiv \epsilon_{kqs} \langle v_q\partial_l\int^t v_s
dt\rangle,\hfil\eqno(21b)$$
$$\hfil D_{ij}\equiv \langle v_i \int^t v_j dt\rangle.\hfil\eqno(21c)$$
The $D_{ij}$ tensor is the usual diffusion tensor.  The $\alpha_{ij}$
tensor is known as the helicity tensor and its trace is the helicity of
the velocity field.  The derivation of eq. (21) assumes an
incompressible velocity field, which is a reasonable approximation
in the context of the motions proposed to drive an accretion disk
dynamo.

Equation (21) is based on the assumption that the back reaction from the
small scale field is negligible and that the large scale field evolution
time scale is long compared to the relevant time scales for $\vec b$
and $\vec v$.  In some circumstances $\vec b$ can have a dramatic
effect on $\vec v$, and consequently on the growth of $\vec B$, and
this can lead to a saturation of the dynamo process.  This can usually
be taken into account by modifying the definition of $\alpha_{ij}$.
A more serious problem is that the second term in eq. (21a) has a
diffusive role, analogous to the (neglected) effects of ohmic
dissipation (Parker 1971; Vainshtein and Ruzmaikin 1971, 1972).
This actually represents a diffusion of the likely
location of large scale magnetic field lines, but in order to
use this `turbulent diffusion' as a substitute for ohmic dissipation
we need to believe that small scale reconnection of field lines
actually removes local sharp gradients in the field rather than
merely moving them around in an unpredictable manner.  Recently
Vainstain and Cattaneo (1992) have pointed out that
if this reconnection is assumed to take place on scales characteristic
of ohmic dissipation then the amount of small scale magnetic field energy
required by the reconnection of large scale field lines is prohibitively
large.  The back reaction from such fields would lead to a saturation
of the dynamo process when the large scale field strength was
almost completely negligible.  The Kraichnan spectrum
(Kraichnan 1965) for MHD turbulence predicts little energy on
such scales and suggests that large scale field lines will be relatively
unbent by a strongly turbulent cascade if the magnetic field energy
and fluid kinetic energy are roughly equal on large scales.
Since large scale fast dynamos do appear to exist this poses
something of a paradox, whose resolution is
not entirely clear.  One possibility for the Sun is that relatively
large scale loops can be created and transported by convective flows, so that
reconnection events consist of the mutual annihilation of large scale,
oppositely oriented field lines (Vainshtein, Parker, and Rosner 1993). This
should also work in
accretion disks where the MSI will constantly create loops of
size $V_A/\Omega$ and differential rotation should help increase
field gradients.  The reader should bear in mind that this plausible
suggestion is not supported, either in the Sun or in accretion disks,
by detailed calculations.

In a differentially rotating disk eq. (21) must be modified
to take into account the background flow of material.  This has
the effect of adding the term $r\partial_r\Omega B_r\hat\theta$ to
the right hand side.  In an accretion disk this term will dominate the
growth of $B_\theta$.  This leads to a cycle in which fluid motions,
acting through the $\alpha_{ij}$ term generate $B_r$ from $B_\theta$.
The differential shearing of the disk then closes the loop, and the
dissipative terms provide the damping.  This is
referred to as an $\alpha-\Omega$ dynamo.
The total dynamo growth rate will be less than $\Omega$ (since the
generation of $B_r$ from $B_\theta$ will tend to be inefficient).
This in turn implies that the dominant large scale field component
will be $B_\theta$.  For a thin accretion disk the only important
spatial gradient in $\vec B$ (or in $\alpha_{ij}$ and $D_{ij}$)
will be in the $\hat z$ direction.  This allows us to produce a
simplified dynamo equation of the form
$$\partial_t B_r=-\partial_z(\alpha_{\theta\theta}B_\theta)+\partial_z
(D_{zz}\partial_z B_r),\hfil\eqno(22a)$$
and
$$\partial_t B_\theta=-{3\over2}\Omega B_r+\partial_z(D_{zz}\partial_z
B_\theta),\hfil\eqno(22b)$$
with $\alpha_{\theta\theta}$ and $D_{zz}$ as defined in eqs. (21).
Note that we have dropped the third term in eq. (21a).  As long
as $\vec B$ is largely azimuthal this term will be very small.
A more exact treatment would include the $\alpha_{\theta r}B_r$ term
in eq. (22a), but such a term cannot, by itself, drive a dynamo since
a 2 dimensional dynamo is impossible (Zel`dovich 1956).
The requirement that $\vec\nabla\vec B=0$ implies that $B_z$ is
of order $(H/r)B_r$ for the internally generated field in a disk,
which makes it dynamically negligible.

{}From eq. (22) we can see that if a growing mode exists its growth
rate will be approximately
$$\hfil\Gamma_{dynamo}\sim \left({\alpha_{\theta\theta}\Omega\over
H}\right)^{1/2}.\hfil\eqno(23)$$
Of course, this estimate is only realistic if we can neglect the damping
terms.  The existence of a growing mode requires that
$$\hfil\left({\alpha_{\theta\theta}H^3\Omega\over
D_{zz}^2}\right)>1.\hfil\eqno(24)$$
where $\alpha_{\theta\theta}$ and $D_{zz}$ are suitably averaged over the
vertical structure of the disk.  Equation (24) assumes that the vertical
scale height of all the relevant physical quantities is $H$.  More
realistically,
one has to solve the appropriate vertical structure equations with a given
model of the small scale velocities.

So far we have not addressed the question of the velocity field that drives
$\alpha_{\theta\theta}$, although we have mentioned several candidates
in previous sections.  We will go through the list of candidate processes
starting with those least likely to drive a dynamo successfully.
We will defer a discussion of the wave driven dynamo until the next section.
First on this list is the Parker
instability.  The unstable modes of magnetic buoyancy will have
$k_z\sim H^{-1}$, $(m/r)\sim H^{-1}$, and $k_r\sim \Omega/V_A$.  Turbulent
mixing due to MSI implies a coherence time for these modes of $\Omega^{-1}$.
The vertical acceleration will be of order $V_A^2/H$ so the vertical
velocity after one coherence time will be $V_A^2/c_s$.  Since the flow
is approximately incompressible this implies $v_r\sim (k_z/k_r)v_z$ or
$v_r\sim (V_A/c_s)^3c_s$.  It follows that
$$\hfil \alpha_{\theta\theta}\sim \left({V_A\over c_s}\right)^5
c_s.\hfil\eqno(25)$$
The minimum reasonable value for $D_{zz}$ is the one given by the MSI, or
$V_A^2/\Omega$.  Comparing eqs. (24) and (25) we can immediately see
that magnetic buoyancy cannot drive a successful dynamo when $H\ll r$.

The next candidate mechanism is the MSI.  There is almost no work on the
ability
of the azimuthal MSI to drive a dynamo, although Zhang {\elevenit et al.}
(1992)
have shown that the vertical MSI can do so.  This result is interesting
from the point of view of supplying a seed field for some other dynamo
process, but by itself appears to generate an azimuthal field whose
final strength depends on the strength of the applied external field.
As far as the azimuthal MSI goes we can only make some tentative and
qualitative comments.  We start by noting that eqs. (15) and (16)
contain neither vertical structure nor local buoyancy.  Although this
should be adequate for modes with rapid growth rates and small vertical
wavelengths, it leaves us with a set of relationships that give no
indication of any loss of local symmetry that would make
$\alpha_{\theta\theta}$
nonzero.  If we explicitly calculate $\alpha_{\theta\theta}$ we obtain
$$\hfill\alpha_{\theta\theta}={-k_z\over k_*^2}\left({m\over
r}\right)^2\Omega\langle\left({1\over\bar\omega^*}\left(1+
{4\omega_A^2\over\tilde\omega^{*2}}\right)\right)|v_r|^2\rangle\tau_{corr}
\hfil\eqno(26)$$
where $\tau_{corr}$ is the correlation time for the mode, which in
the saturated state should be roughly the inverse of the growth
rate.  This expression should be zero (after averaging over a set of
similar modes) for two reasons.  First, the real contributions come from
terms that are odd functions of the distance to the corotation point,
and will integrate to zero after radial averaging.  Second,
there is no obvious reason why one sign of $k_z$ should be preferred.
This argument has the loophole that it is implicitly weakly nonlinear,
whereas the saturated state of the instability will leave these modes
in a strongly nonlinear state.  Nevertheless, the basic lack of a means of
generating the necessary asymmetries remains.
We might instead appeal to modes with $k_zH\sim 1$, which have
the possibility of significant contributions from buoyancy effects and
vertical structure terms, but we can see from eq. (17) that
the maximum growth rate on these scales is $\sim (k_z/k)\Omega\sim V_A/H$.
Such modes will have $k_z\sim (m/r)\sim H^{-1}$ and $k_r\sim \Omega/V_A$.
They will be suppressed for the same reason that magnetic buoyancy modes
are suppressed, and by an identical argument can be shown to be incapable
of driving a dynamo.  Of course, if either of these processes could be
shown to work in spite of these difficulties then one would have a theory
of angular momentum transport which would be a true $\alpha$ model, in
the sense that $\alpha$ would be a constant of order unity describing
a purely local process of dissipation.

It turns out that convection is a particularly interesting candidate
for driving a dynamo, a point first raised by Galeev, Rosner and Vaiana (1979).
When convection is weak, with a growth rate $\Gamma_c\ll\Omega$  we have
seen that it is characterized by
$k_z\sim (m/r)\sim H^{-1}$ and $k_r\sim (\Omega/\Gamma_c)H^{-1}$ with
$v_z$ and $v_\theta$ of order $H\Gamma_c$ and $v_r$ smaller by a factor
of $\Gamma_c/\Omega$.
When the MSI mixes a convective cell faster than $\Gamma_c$ then
convection will be disrupted and, like the Parker instability, convection
becomes a poor prospect for the source of a successful dynamo.  On the
other hand, when
$$\hfil k_r^2 V_A^2 \Omega^{-1}\sim \left({V_A\over
H\Gamma_c}\right)^2<\Gamma_c\hfil\eqno(27)$$
then convection will proceed unimpeded.  In this case, vertical mixing occurs
at a rate of $\Gamma_c$ so that $D_{zz}\sim H^2\Gamma_c$.  The correlation
time of the convective flow is approximately $\Gamma_c^{-1}$ so
$$\alpha_{\theta\theta}\sim \left({\Gamma_c\over
\Omega}\right)^2c_s.\hfil\eqno(28)$$
We see from these results and from eq. (23) that convection is
marginally capable of driving a dynamo.  Even if it is unable to drive
a large scale dynamo it is clear that the convective cells will tend to
wind up the magnetic field lines of a disordered field.  The local field
strength will consequently grow until the MSI associated with it is strong
enough to disrupt the convective cells.  When eq. (27) is replaced
by an equality then the angular momentum transfer associated with convection
will be approximately equal to that associated with MSI.  The difference
is that the  MSI will clearly induce an outward transfer of angular momentum,
unlike purely
hydrodynamic convection whose effect is uncertain.  It seems possible
that sufficiently strong convection in a {\elevenit conducting} medium
transports angular momentum outward with a dimensionless `viscosity'
$\alpha\sim (\Gamma_c/\Omega)^3$.
In addition, turbulent diffusion and magnetic buoyancy would tend to
spread the magnetic field generated within a zone of strong convection
until such fields were found for all $z$ at any radius where convection
occurs.  Such fields would, of course, be subject to the MSI and promote
outward angular momentum transport.
\vskip .6cm
\line{\elevenbf 4. The Internal Wave Driven Dynamo\hfil}
\vskip .4cm
At this point we have a good candidate for the local driving mechanism
behind angular momentum transport (the MSI), but the best dynamo
candidate discussed above is convection, which will not always exist, or
necessarily
sustain itself self-consistently, and cannot be guaranteed to drive
angular momentum outward instead of inward.  Clearly we need
an alternative.
One attractive possibility is that the internal wave field
discussed in section 2.2. might drive a dynamo.  A similar point was originally
raised
by Pudritz (1981a,b) who envisioned the waves as arising spontaneously
from some local instability in the disk.  This process does not seem to
occur, but the global excitation of such modes can stand in its place.
Our results here are taken from Vishniac, Jin and Diamond (1990) and
differ from Pudritz's treatment not only in the source and amplitude of
the wave field, but also in the details of the derivation of
$\alpha_{\theta\theta}$ from such waves.
\vskip .6cm
\line{\elevenit 4.1. Internal Waves as a Source of Helicity\hfil}
\vskip .4cm
We have already argued
that the dominant internal wave modes have $k_z\sim H^{-1}$ and $\bar\omega$
largely
real.  Starting from eq. (2) we find that
$$\hfil\alpha_{\theta\theta}\approx S^2\langle {k_r {m\over
r}\bar\omega^2\over(\Omega^2-\bar\omega^2)(\bar\omega^2-N^2)}\partial_z\chi^*
\chi\rangle\tau_{corr}.\hfil\eqno(29)$$
We have assumed $m$ small, but not not necessarily $\pm 1$ since nonlinear
interactions will generate other values of $m$.  This expression will be
nonzero
for any given mode, but unless the sign of $mk_r$ is fixed the contributions
from different modes will cancel out.  Fortunately the sign of $mk_r$ is fixed
by the expectation that the wave field will be dominated by modes that are
undergoing linear amplification.  We can see from eq. (7) that internal
waves have a radial group velocity with the opposite sign of their phase
velocity, {\elevenit i.e.} $sign[V_{gr}]=sign[\bar\omega/k_r]$.  The
requirement that internal
waves are undergoing amplification
is that they be headed for reflection, {\elevenit i.e.} that $\bar\omega^2$
increase in the
direction of propagation, implying that $sign[V_{gr}\partial\bar\omega^2]>0$.
This in turn tells us that
$$\hfil\bar\omega k_r\partial\bar\omega^2=-{3m\over r}\Omega\bar\omega^2
k_r>0.\hfil\eqno(30)$$
We conclude that $mk_r<0$ for modes undergoing linear amplification.
The existence of strong differential rotation in the disk fixes
the sign of $mk_r$ of the dominant wave modes, which in turns determines the
sign
of their contribution to $\alpha_{\theta\theta}$.

It is useful to rewrite eq. (29) in terms of the mean square velocity contained
in
modes with a given $m$ and $\bar\omega$.  Unless $\bar\omega$ is very close to
$\Omega$
we have $\langle v^2\rangle\sim (S\chi k_z/\bar\omega)^2$ so that
$$\hfil\alpha_{\theta\theta}\sim \langle v^2\rangle
\left({\bar\omega^2\over\Omega}\right)^2k_r {m\over
r}H\tau_{corr},\hfil\eqno(30)$$
where we have neglected $N^2$.  Including it would confine the waves, and the
helicity, to the region
around the midplane where $\bar\omega^2>N^2$, but within that region properly
accounting for
$N^2$ will introduce a correction of order unity.  Vishniac and Diamond (1992)
have shown that
including the confinement of the waves does not affect the vertically averaged
angular momentum transport significantly so we will ignore this point in what
follows.
For realistic vertical structure $\alpha_{\theta\theta}$ will be an odd
function of the distance to the midplane
of the disk.  For strongly turbulent flows $\tau_{corr}$ is the eddy turnover
time.  For waves
it is $\Re (i\bar\omega)^{-1}$.  This is zero for waves with exactly real
frequencies.  Purely periodic
motions do not result in any net transport.  For finite amplitude waves,
however,
nonlinear interactions between the waves
give rise to a small imaginary frequency, of order $\tau_{decorr}^{-1}$.
Consequently,
$\tau_{corr}\approx (\bar\omega^2\tau_{decorr})^{-1}$.  We conclude that
$$\hfil\alpha_{\theta\theta}\sim \langle
v^2\rangle\tau_{decorr}^{-1}\left({1\over\Omega}^2\right)k_r {m\over
r}H.\hfil\eqno(31)$$
If we imagine that the energy from the fundamental, global wave mode is passed
through a weakly turbulent
cascade down to small scales, then within that cascade we expect $\langle
v^2\rangle \tau_{decorr}^{-1}$ integrated
over a logarithmic interval to be roughly constant.  To the extent that further
linear amplification
can take place within this cascade $\langle v^2\rangle\tau_{decorr}^{-1}$ might
even increase.
Consequently eq. (31) suggests that the helicity will be
dominated by modes with large $m$ and $k_r$.
Of course, this argument fails if the sign of $mk_r$ is not preserved in the
cascade, but preliminary
calculations (Huang 1992) suggest that it is preserved.
\vskip .6cm
\line{\elevenit 4.2. The Internal Wave Spectrum\hfil}
\vskip .4cm
Given that $\alpha_{\theta\theta}$ is likely to be dominated by small scale
waves, generated from
the nonlinear dissipation of the global modes, we need some understanding of
the course of
nonlinear dissipation.  Here we appeal to a simple model for the cascade,
discussed
in Vishniac, Jin, and Diamond (1990) and Vishniac and Diamond (1992).  Previous
work on the dissipation of deep ocean waves (McComas and Bretherton 1977) has
shown
that there are three basic processes mediating the nonlinear transfer of energy
in
a weakly turbulent cascade: elastic scattering, induced diffusion, and
parametric subharmonic
instability.  The first two processes are unimportant in this context.  The
third
involves the generation of waves from a wave with roughly twice the radial
wavenumber
and half the frequency of the generated waves.  The vertical wavenumber changes
only
slightly in this process (remember that this is a continuum model!).  The
analog for
an accretion disk is that the global modes tend to generate local modes with
$k_z\sim H^{-1}$
but steadily decreasing $\bar\omega$ and increasing $k_r$.  This process will
eventually
terminate in the eruption of strong turbulence at some suitably small scale.
This
model makes no firm prediction as to the sign of $mk_r$ in the cascade or the
magnitude of $m$
as a function of $k_r$.  Numerical work by Huang (1992) based on the weakly
resonant approximation
(Phillips 1960, 1961) supports this basic picture and also suggests that $m$
increases slowly
in the cascade while the sign of $mk_r$ is preserved.  In what follows we will
take
$\langle v^2\rangle \tau_{decorr}^{-1}$ as a constant in the cascade.  This can
be justified in detail by comparing $\tau_{decorr}$ and the time required for
linear
amplification.  This implies that $\langle v^2\rangle\propto \bar\omega^{1/2}$
and
$\tau_{decorr}\propto \bar\omega^{1/2}$.  Since $\langle v^2\rangle\sim
(H/r)c_s^2$ for the
global modes which lie at the top of the cascade we have
$$\hfil\langle v^2\rangle_{\bar\omega}\sim {H\over
r}c_s^2\left({\bar\omega\over\Omega}\right)^{1/2},\hfil\eqno(32a)$$
and
$$\hfil\tau_{decorr,\bar\omega}^{-1}\sim {H\over
r}\left({\Omega\over\bar\omega}\right)^{1/2}\Omega.\hfil\eqno(32b)$$
We will also assume $m^2\propto\bar\omega^{-1}$, but our results change only
slightly if the increase in $m$
is more gradual.
We obtain from eqs. (31) and (32)
$$\hfil \alpha_{\theta\theta}\sim \left({H\over
r}\right)^3\left({\Omega\over\bar\omega}\right)^{3/2}c_s.\hfil\eqno(33)$$
All of the previous work cited is based on a random phase approximation, so our
use of it here implies
a neglect of soliton formation in the ingoing internal waves.

In an unmagnetized disk the cascade will extend down to scales
where\hfill\break $\langle v^2\rangle^{1/2}>H\bar\omega$.  On
smaller scales the fluid will show strong, anisotropic turbulence.  In a
magnetized disk these results will
be valid only if $\bar\omega>(m/r)V_A$ and if the turbulence due to MSI
dissipates the waves more slowly
than $\tau_{decorr}^{-1}$.  Below this scale the fluid motions will be rapidly
dissipated by the small scale
turbulence induced by the MIS.  The former condition is necessary for our
treatment of internal waves as purely
hydrodynamic motions to be valid.  It is also easy to satisfy.  The latter
condition
is important in providing a natural saturation mechanism for the dynamo.  From
eq. (32b) we can see that it
is equivalent to
$$\hfil k_r^2V_A^2<\Omega^2\left({H\over
r}\right)\left({\Omega\over\bar\omega}\right)^{1/2},\hfil\eqno(34a)$$
or
$$\bar\omega>\left({r\over H}\right)^{2/3}\left({V_A\over
c_s}\right)^{4/3}\Omega.\hfil\eqno(34b)$$
As $V_A$ increases this limit will come to define the lower end of the weakly
turbulent wave cascade.  Since
$\alpha_{\theta\theta}$ is dominated by waves at this end of the spectrum the
increase in $V_A$ will have the
effect of turning down the dynamo growth rate and increasing the turbulent
dissipation rate until they balance.
\vskip .6cm
\line{\elevenit 4.3. Scaling Laws for the Internal Wave Driven Dynamo\hfil}
\vskip .4cm
We are now in a position to derive the basic scaling laws that describe the
internal
wave driven dynamo model.  When the dynamo is saturated, the dynamo growth
rate,
given in eq. (23), must be balanced by turbulent dissipation.  Since the MSI
induces roughly isotropic turbulence this implies that
$$\hfil \Gamma_{dynamo}\sim \left({V_A\over
c_s}\right)^2\Omega.\hfill\eqno(35)$$
{}From eqs. (23) and (33) we see that $\Gamma_{dynamo}$ is dominated by the
contribution from wave modes with the smallest values of $\bar\omega$ that
still
satisfy the inequality given in eq. (34b).  Combining these results we
find that in a stationary disk, with a saturated dynamo,
$$\hfil V_A\sim c_s\left({H\over r}\right)^{2/3},\hfil\eqno(36a)$$
$$\hfil \alpha_{\theta\theta}\sim c_s\left({H\over
r}\right)^{8/3},\hfil\eqno(36b)$$
$$\hfil \alpha\sim \left({V_A\over c_s}\right)^2\sim\left({H\over
r}\right)^{4/3},\hfil\eqno(36c)$$
and
$$\bar\omega_c\sim \left({H\over r}\right)^{2/9}\Omega,\hfil\eqno(36d)$$
where $\bar\omega_c$ is the frequency for which eq. (34b) is an equality.
We see that this frequency is rather insensitive to the shape of the accretion
disk,
and will be small, relative to $\Omega$, only for extremely flattened systems.
We
have also neglected corrections due to the vertical structure of the disk and
its
ability to confine waves with small $\bar\omega$ to regions very close to the
midplane.  An approximate correction to $\alpha$ due to this effect can be
found
in Vishniac and Diamond (1992).  It changes eq. (36c) by a factor of
only $(H/\Delta z_c)^{1/6}$, where $\Delta z_c$ is the height of the region
where waves of frequency $\bar\omega_c$ can propagate.  For a nearly adiabatic
disk this is a factor of $\sim 1$, but even for an isothermal disk this factor
is only $(H/r)^{-2/15}$.

The model of the turbulent spectrum that we have used to arrive at these
results is
somewhat crude. Still, we can see that the exact cutoff frequency and dynamo
growth rate
are not terribly sensitive functions of $(H/r)$ and will not be very different
if we
had assumed $m$ constant (or increasing as $\bar\omega^{-1}$) within the
turbulent
cascade.  Taking a somewhat broader view, we can see that no model of the
turbulent
cascade, even one incorporating soliton effects, will give $\alpha$ outside the
range
of $(H/r)^{1-1.5}$.  At the upper limit for the exponent we have the
contribution due
to the global modes alone.  At the lower limit the MSI turbulence will start to
interfere with the propagation of those modes, and thereby saturate the entire
process.

At this point it is useful to summarize the basic physics of the internal wave
driven
dynamo.  The flowchart below illustrates the way the various
elements of this model interact to drive mass inward in an accretion disk.
Properly speaking the scaling laws we have quoted here apply only to a
stationary system.
For disk evolution rates in excess of $\Gamma_{dynamo}$ the arrows on this
flowchart
must be replaced by time dependent equations.  For evolution rates in excess of
the wave
propagation rate, $(H/r)\Omega$, deviations from these stationary estimates can
be large.
\vskip 0.2cm
\centerline{EXCITATION OF INTERNAL WAVES}
\vskip 0.3cm
\line{\hskip 7cm $\Big\downarrow$\ (+ Linear Amplification)\hfill}
\vskip 0.3cm
\line{\hskip 7cm $\Big\downarrow$\ (+ Nonlinear Wave Interactions)\hfill}
\vskip 0.3cm
\centerline{SATURATED WAVE SPECTRUM}
\vskip 0.3cm
\line{\hskip 7cm $\Big\downarrow$\ (+ Shearing)\hfill}
\vskip 0.3cm
\centerline{$\alpha-\Omega$ DYNAMO}
\vskip 0.3cm
\line{\hskip 7cm $\Big\downarrow$\hfill}
\vskip 0.3cm
\centerline{GROWTH OF $B_\theta$}
\vskip 0.3cm
\line{\hskip 7cm $\Big\downarrow$\ (+ Magnetic Shearing Instability)\hfill}
\vskip 0.3cm
\centerline{SMALL SCALE TURBULENCE}
\vskip 0.3cm
\line{\hskip 7cm $\Big\downarrow$\hfill}
\nointerlineskip
\moveright 1.13cm \vbox{\hrule width11.2cm}
\nointerlineskip
\line{\hskip 1cm $\Big\downarrow$\hskip 3.5cm $\Big\downarrow$\hskip 3.5cm
$\Big\downarrow$\hskip 3.5cm $\Big\downarrow$\hfill}
\vskip 0.3cm
\line{SATURATION\hskip 1cm TURBULENT\hskip 1cm TRUNCATION\hskip 1cm
SUPPRESSION\hfill}
\line{\hskip 1cm OF $B_{\theta}$\hskip 1.6cm TRANSPORT\hskip 1.1cm OF
INTERNAL\hskip 0.9cm OF
PARKER\hfill}
\line{\hskip 7.2cm WAVE SPECTRUM\hskip 0.5cm INSTABILITY\hfill}
\vskip .6cm
\line{\elevenit 4.4. Phenomenology of the Internal Wave Driven Dynamo\hfil}
\vskip .4cm
We have seen that the internal wave driven dynamo gives a result for the
angular momentum
transport in a stationary disk which is apparently consistent with
phenomenological models.
This is actually not certain since this mechanism is non-local and may show
significant
deviations from a local model with $\alpha\sim (H/r)^{4/3}$.  At this point not
much
is known about where such deviations might show up.  Here we comment on a few
phenomenological
aspects of this model that can be derived from qualitative considerations.

It is well known that $\alpha$ model accretion disks dominated by electron
scattering whose main
pressure support is due to radiation are subject to thermal and viscous
instabilities
(Pringle, Rees, and Pacholzcyk 1973; Lightman and Eardley 1974).
Lightman and Eardley also noted that these instabilities
can be cured by excluding the radiation pressure from the definition of
viscosity, but this
suggestion has no obvious physical basis. It is certainly not a consequence of
the internal
wave driven dynamo model.  Fortunately, it turns out that these instabilities
are largely,
though not entirely, suppressed in this model.  In order to see this we need to
review the basis
of these instabilities.  The rate at which local thermal equilibrium is
established in
a disk, $\tau_t^{-1}$, is the local flux divided by the vertical integral of
the thermal energy.  The
rate at which a stationary surface density distribution is established,
$\tau_v^{-1}$ is the mass
transfer rate divided by $\Sigma r^2$.  For a stationary disk dominated by
electron scattering
and radiation pressure we have
$$\hfill H\sim {F_\gamma\sigma_T\over c\mu}\sim {P_\gamma\over \Sigma
\Omega^2},\hfill\eqno(37a)$$
$$\hfill\tau_v^{-1}\sim {\dot M\over\Sigma r^2}\sim \alpha\left({H\over
r}\right)^2\Omega,\hfill\eqno(37b)$$
and
$$\hfill\tau_t^{-1}\sim {\dot M\Omega^2\over P_\gamma H}\sim
\alpha\Omega,\hfill\eqno(37c)$$
where $F_\gamma$ is the photon flux from the disk and $P_\gamma$ is the photon
pressure near the
midplane.  The first condition is the condition for
hydrostatic equilibrium and is established at a rate of $\sim\Omega$.  Consider
first a
perturbation to the structure of the disk in which some annulus gains an extra
bit of thermal
energy.  The heat generation rate is just $\dot M\Omega^2$ or $\alpha\Sigma
H^2\Omega^3$.
On thermal time scales $\Sigma$ is a constant, so the heating rate grows as
$\alpha H^2$, where
$H$ is proportional to $P_\gamma$.  On the other hand, the cooling rate is
$$\hfill F_\gamma\sim {P_\gamma c\over n_e\sigma_T H}\sim {P_\gamma
c\mu\over\Sigma}\hfill\eqno(38)$$
where $\mu$ is the average mass per electron.  This is proportional to $H$ so
the heating rate
grows more quickly than the cooling rate and the disk undergoes a phase of
runaway heating.
Conversely, if we had considered an initial drop in the local value of
$P_\gamma$ the disk
would, locally, have undergone runaway cooling.  This instability can be cured
if $\alpha$
is proportional to $H^n$, with $n<-1$.  At first glance it would appear that
the internal
wave driven dynamo model, and any model designed to reproduce the phenomenology
of cataclysmic
variable disks, can only exacerbate this problem.  Moreover, since the annulus
of the perturbed
region has not explicitly entered our discussion, we can expect this difficulty
to crop up on
all scales larger than $H$.

These disks are also viscously unstable, as we can show by considering longer
time scale variations
and assuming that thermal equilibrium is somehow maintained.  If $F_\gamma\sim
\dot M\Omega^2$
then
$$\hfill H\sim {\dot M\sigma_T\over c\mu}.\hfill\eqno(39)$$
Then remembering that $\dot M\sim\alpha\Sigma H^2\Omega$ we see that $\dot
M\propto (\Sigma\alpha)^{-1}$.
For a constant $\alpha$ we note that overdense regions in the disk have a
smaller $\dot M$ and
will become increasingly dense and cold, while underdense regions in the disk
have a larger
$\dot M$ and will become increasingly diffuse and hot.  Again, this instability
can be evaded
if $\alpha\propto H^n$ with $n<-1$.  It is not clear what the consequence of
these instabilities
might be.  Apparently, they make it impossible to construct a locally sensible
theory of
accretion disks in this limit.

It turns out that in the internal wave driven dynamo model these instabilities
are completely
suppressed for annuli with widths much smaller than the disk radius.  We start
by considering
very short time scales, shorter than $\Gamma_{dynamo}^{-1}$.  From flux
conservation we see that
$B_\theta\propto H^{-1}$. Consequently $\alpha\propto (V_A/c_s)^2\propto
(H^3\Sigma)^{-1}$ or
$\alpha\propto H^{-3}$.  We see immediately that this suppresses instabilities
on such short
time scales.  Since $\tau_t^{-1}\sim \Gamma_{dynamo}$, {\elevenit i.e.} the
critical
time scale is
just long enough to force us to consider the evolution of the dynamo equations,
this estimate
does not answer the general question
of the stability of narrow annuli in radiation pressure and electron scattering
dominated
disks.  Suppose we consider longer time scales.  As long as the annulus is
narrow compared
to $r$ the internal wave energy within the annulus is fixed by the condition
that the
flux supplied from larger $r$ is constant.  In an annulus that is thicker than
its upstream
neighbor the waves will spread out and speed up.  We can express the result in
terms of $R$,
the ratio of the mean square perturbed velocities contained in $|m|=1$ waves
relative to
the normal saturated value.  Since $\Sigma\langle v^2\rangle c_s$ is constant
we have
$$\hfill R\equiv \left({r\over H}\right)\left({\langle v^2\rangle\over
c_s^2}\right)\propto {1\over\Sigma c_s^3}.\hfill\eqno(40)$$
Of course, a change in $R$ affects the nonlinear energy transfer rate as $R^2$.
 Tracking the effects
of a change in $R$ through our previous results we find that $\bar\omega_c$ is
unaffected but
that $\alpha$ is proportional to $R$.  In other words
$$\hfill\alpha\sim \left({H\over r}\right)^{4/3}R.\hfill\eqno(41)$$
Combining eqs. (40) and (41) we see that on thermal time
scales $\alpha\propto H^{-5/3}\Sigma^{-1}$.
Short radial wavelength thermal instabilities are suppressed.  On long radial
scales the internal
wave flux can grow (and shrink) to track the saturation value of $\langle
v^2\rangle$, implying
the {\elevenit possibility} of thermal instabilities on such scales.  Even if
these occur they
do not affect our ability to construct locally sensible models of an accretion
disk.  Such
long radial scale instabilities
would imply that the disk luminosity would vary significantly on time scales
$\sim (H/r)^{-4/3}\Omega^{-1}$.
These variations could not be arbitrarily large since once $\delta H/H$ was of
order unity across
the unstable annulus the waves would no longer be able to grow, or decay, fast
enough to remain
near the normal saturation limit.

If we now consider viscous time scales, so that $\Sigma$ is no longer held
constant, we find
that $\dot M$ is not a function of $\Sigma$ at all.  The internal wave driven
dynamo model
for radiation pressure and electron scattering dominated disks is neutrally
stable against
short wavelength viscous instabilities.  Low levels of absorption and gas
pressure will
stabilize these modes, but on longer radial scales we expect that viscous
instabilities
will be much harder to suppress than thermal instabilities.  We therefore
expect these
disks to show luminosity variations on time scales of
$(H/r)^{-10/3}\Omega^{-1}$.

One point that will be a generic feature of {\elevenit any} model that relies
on the MSI
of an azimuthal magnetic field is that the dominant turbulent eddys are nearly
isotropic.
Consequently, the vertical diffusion coefficient, $V_A^2/\Omega$, is comparable
to the
radial diffusion coefficient and also to $\alpha Hc_s$.  This implies that
there
will always be a minimal level of vertical mixing taking place, even in
convectively
{\elevenit stable} accretion disks.  To see how important this is, we can
compare the
radiative flux per unit area in a stationary disk, $\dot M\Omega^2$, to the
typical
vertical heat transport due to mixing.  The latter is $\sim P D_z/L_s$, where
$L_s$
is the entropy scale height.  The ratio between these two transport terms is
$\sim H/L_s$,
which will usually be of order unity.  In other words, vertical radiative
transport
in accretion disks must always compete with hydrodynamic mixing effects of the
same order.
Of course, in a stably stratified disk the latter will tend to move entropy
{\elevenit inward},
implying a greater temperature difference between the midplane and photosphere
than for
purely radiative models.

Another interesting point about this model is that a fair amount of energy is
available
for coronal heating.  The magnetic field energy per unit area in the disk is
$\sim (H/r)^{4/3}PH$.
This will diffuse into the disk atmosphere at a rate of $\sim
(H/r)^{4/3}\Omega$ so we see
that the fraction of the energy dissipated locally which is available for
magnetic heating
of the corona is $\sim (H/r)^{4/3}$.  On the other hand, the internal wave
energy per unit
area is $\sim(H/r)PH$.  Nonlinear dissipation will dump this energy into
neighboring modes
in phase space at a rate of $\sim (H/r)\Omega$.  Most internal waves in the
disk will only
couple efficiently to other internal waves, but the global modes are not far
removed from
the lowest frequency sound waves and we can expect some fraction of their
energy to
be transferred to sound waves.  Eventually a large fraction of the energy in
sound waves
will end up as shocks in the disk atmosphere giving a heating rate of $\sim
(H/r)^2PH\Omega$
per unit area, or an efficiency for coronal heating that will scale as
$(H/r)^{2/3}$.
Although the scaling law for shock heating is more favorable, the coefficient
is likely
to be smaller since it relies on a series of inefficient intermediate steps.
Unless $H/r$ is
very small both processes are likely to be important in coronal heating.

Finally, we note that we have, by hypothesis,  excluded external magnetic
fields from our
discussion, but there are
certainly circumstances where such fields are likely have a large effect on the
structure
of the disk.  In extreme cases, like intermediate polars, the magnetic field of
the central star
may be large enough to disrupt the disk entirely inside some critical radius
(Ghosh and Lamb 1979).
This should certainly
be the case when the energy density of the imposed field exceeds the kinetic
energy density of
the disk, {\elevenit i.e.} $B_z^2/(4\pi)> \rho (r\Omega)^2/2$.  At smaller
field strengths
the presence of the field can lead to disk instabilities which can increase the
effective value of $\alpha$ within the disk.  As an example, we can consider
the MSI due to an
external magnetic field.  As we have already seen, such a field can give rise
to a turbulent
viscosity $\sim V_A H$, where $V_A$ is the Alfv\'en velocity due to the
external field only.
When $(V_A/c_s)>(H/r)^{4/3}$ this will be the dominant source of angular
momentum transport.
The MSI is an incompressible instability, which operates only when the magnetic
pressure is
smaller than the fluid pressure, but similar effects can arise from
compressional modes
(Kaisig, Tajima, and Lovelace 1992; Tagger, Pellat, and Coroniti 1992;
Tagger, Henriksen, Pellat, and Sygnet 1990) for an
arbitrarily cold disk, {\elevenit i.e.} a low $\beta$.  In this case the
effects
of the disk fluid pressure become small and on dimensional grounds we expect
the the effective
turbulent viscosity to
scale as $V_A^2/\Omega$.  We see that the crossover point between the two
mechanisms
is, as expected, roughly where the magnetic field pressure from the external
field
and the fluid pressure within the disk are comparable.  Both mechanisms are
strongly
dependent on the strength of the external field.  They may be important close
to the
central object, or close to the inner boundary of the disk, but they will
quickly
become unimportant as we move to larger radii.  In the region where they are
important
these effects must compete with the tendency of the central object to exert a
torque on
the disk through the stretching of its magnetosphere (Ghosh and Lamb 1979;
Mauche, Miller, Raymond
and Lamb 1990).  The torque exerted
in this way is $B_zB_\theta r/\Sigma$.  Assuming that $B_\theta\sim B_z
\Omega_*/\Omega$
this will be a large, even overwhelming, effect whenever the vertical MSI or
compressional mode
effects are strong enough to compete with the turbulent transport generated
within the
disk by purely internal processes.
\vskip .6cm
\line{\elevenbf 5. Conclusions\hfil}
\vskip .4cm
More than 20 years after the seminal theoretical paper by Shakura and Sunyaev
the transport
of angular momentum in accretion disks remains a contentious topic.  In spite
of
this several points have emerged that are likely to be important in
any future theory.  First, the direction of angular momentum transport is not
guaranteed by appeals to turbulent mixing.  This makes models based on
turbulent convection
distinctly unpromising.  Second, magnetic shearing instabilities {\it do}
guarantee
an outward transport of angular momentum and are unavoidable given
a magnetic field.  We believe that this instability will play a critical role
in future
theories of ionized disks.  The implications for cold disks are less clear.
Third, this makes an understanding of disk dynamos absolutely critical.
Fourth, the tendency for successful disk models to include scaling laws of the
form $\alpha\propto (H/r)^n$ argues that a purely local process, {\elevenit
e.g.} a dynamo
driven by the MSI process or by magnetic buoyancy is unlikely to work.  We
expect
that there must be a nonlocal trigger for the dynamo, {\elevenit e.g.} global
waves of some kind.  Finally, if global
waves are important in any way then the interaction between a three dimensional
disk and its environment has to be understood quantitatively.

Here we have presented a model for a dynamo in accretion disks based on
the action of internal waves propagating from the outer edge of the disk.
As the magnetic field grows it gives rise to shearing instabilities which
transport angular momentum outward and magnetic flux vertically.
These instabilities will eventually saturate the dynamo when $\beta$
is still quite large.  The resulting angular momentum transport is
consistent with observations, although the prediction is rather
approximate and is {\elevenit not} strictly equivalent to a local viscosity
in any case.  From the point of view of dynamo theory the interesting feature
here is a dynamo model which is nonturbulent, in the sense that the
turbulent motions present are not responsible for driving the large scale
magnetic field.
{}From the point of view of accretion disk theory the interest of this model
is that it provides a way of understanding angular momentum transport in
accretion disks without resort to phenomenological considerations.

The authors are happy to acknowledge useful contributions from
Min Huang, Liping Jin, Shan Luo, Ptolemy Schwartz and Wenli Zhang.
This work has been supported, in part, by NASA through contract NAGW 2048.
\vskip .6cm
\line{\elevenbf References\hfil}
\vskip .4cm
\line{Adams, F.C., Ruden, S.P. and Shu, F. 1989, {\elevenit ApJ}\ {\elevenbf
347}, 959.\hfill}
\line{Balbus, S. and Hawley, J. 1991, {\elevenit ApJ}\ {\elevenbf 376},
214.\hfill}
\line{Balbus, S. and Hawley, J. 1992, {\elevenit ApJ}\ {\elevenbf 400},
610.\hfill}
\line{Blaes, O.M. 1987, {\elevenit MNRAS}\ {\elevenbf 227}, 975.\hfill}
\line{Cabot, W. Canuto, V.M., Hubickyj, O., and Pollack, J.B. 1987a, {\elevenit
Icarus}\ {\elevenbf 69}, 387.\hfill}
\line{Cabot, W. Canuto, V.M., Hubickyj, O., and Pollack, J.B. 1987b, {\elevenit
Icarus}\ {\elevenbf 69}, 423.\hfill}
\line{Cameron, A.G.W. 1969, in {\elevenit Meteorite Research}, ed.\hfill}
\line{\quad P.M. Millman (Springer: New York), 14.\hfill}
\line{Cannizzo, J.K., Wheeler, J.C., and Polidan, R.S. 1986, {\elevenit ApJ}\
{\elevenbf 330}, 327.\hfill}
\line{Chandrasekhar, S.  1961, {\elevenit Hydrodynamic and Hydromagnetic
Stability},\hfill}
\line{\quad(Oxford University Press: London).\hfill}
\line{Coroniti, F.V. 1981, {\elevenit ApJ}\ {\elevenbf 244}, 587.\hfill}
\line{Donner, K.J. 1979, Ph.D. Thesis, Cambridge University\hfill}
\line{Drury, L.O'C. 1980, {\elevenit MNRAS}\ {\elevenbf 193}, 337.\hfill}
\line{Drury, L.O'C. 1985, {\elevenit MNRAS}\ {\elevenbf 217}, 821.\hfill}
\line{Dubrulle, B. 1992, {\elevenit A\&A} in press.\hfill}
\line{Dubrulle, B. and Knobloch, E. 1992, {\elevenit A\&A}\ {\elevenbf 256},
673.\hfill}
\line{Dubrulle, B. and Valdettaro, L. 1992, {\elevenit A\&A}\ {\elevenbf 263},
387.\hfill}
\line{Dubrulle, B. and Zahn, J.-P. 1991, {\elevenit J. Fluid Mech.}\ {\elevenbf
231}, 561.\hfill}
\line{Eardley, D.M., and Lightman, A.P. 1975, {\elevenit ApJ}\ {\elevenbf 200},
181.\hfill}
\line{Galeev, A.A., Rosner, R. and Vaiana, G.S. 1979, {\elevenbf ApJ}\
{\elevenbf 229}, 318.\hfill}
\line{Ghosh, P. and Lamb, F. K. 1979, {\elevenit ApJ}, {\elevenbf 234},
296.\hfill}
\line{Glatzel, W. 1986, {\elevenit MNRAS}\ {\elevenbf 225}, 227.\hfill}
\line{Glatzel, W. 1988, {\elevenit MNRAS}\ {\elevenbf 231}, 795.\hfill}
\line{Goldreich, P., and Narayan, R. 1985, {\elevenit MNRAS}\ {\elevenbf 213},
7p.\hfill}
\line{Goldreich, P., Goodman, J. and Narayan, R. 1986, {\elevenit MNRAS}\
{\elevenbf 221}, 339.\hfill}
\line{Goodman, J.  1992, {\elevenit ApJ}, in press.\hfill}
\line{Hanawa, T. 1988a, {\elevenit A\&A}\ {\elevenbf 196}, 152.\hfill}
\line{Hanawa, T. 1988b, {\elevenit A\&A}\ {\elevenbf 206}, 1.\hfill}
\line{Hawley, J. and Balbus, S. 1991, {\elevenit ApJ}\ {\elevenbf 376},
223.\hfill}
\line{Hawley, J. and Balbus, S. 1992a, {\elevenit ApJ}\ {\elevenbf 400},
595.\hfill}
\line{Hawley, J. and Balbus, S. 1992b, {\elevenit BAAS}\ {\elevenbf 24},
1234.\hfill}
\line{Huang, M. 1992, Ph.D. Thesis, The University of Texas\hfill}
\line{Huang, M. and Wheeler, J.C. 1989, {\elevenit ApJ}\ {\elevenbf 343},
229.\hfill}
\line{Ichimaru, S. 1977, {\elevenit ApJ}\ {\elevenbf 214}, 840.\hfill}
\line{Iroshnikov, P. 1980, {\elevenit Sov. Astr.}\ {\elevenbf 24}, 565.\hfill}
\line{Kaisig, M. 1989, {\elevenit A\&A}\ {\elevenbf 218}, 89.\hfill}
\line{Kaisig, M., Tajima, T. and Lovelace, R.V.E. 1992, {\elevenit ApJ}\
{\elevenbf 386}, 83.\hfill}
\line{Korycansky, D. 1992, {\elevenit ApJ}\ {\elevenbf 399}, 176.\hfill}
\line{Kraichnan, R.H. 1965, {\elevenit Phys. Fluids}\ {\elevenbf 8},
1385.\hfill}
\line{Larson, R.B. 1990, {\elevenit MNRAS}\ {\elevenbf 243}, 588.\hfill}
\line{Lerner, J. and  Knobloch, E., 1988, {\elevenit J. Fluid Mech.}\
{\elevenbf 189}, 117.\hfill}
\line{Lightman, A.P. and Eardley, D.M. 1974, {\elevenit ApJ Lett.}\ {\elevenbf
187}, L1.\hfill}
\line{Lin, D.N.C. 1981, {\elevenit ApJ}\ {\elevenbf 246}, 972.\hfill}
\line{Lin and Papaloizou, 1980, {\elevenit MNRAS}\ {\elevenbf 191}, 37.\hfill}
\line{Lin, D.N.C., Papaloizou, J. and Faulkner, J. 1985, {\elevenit MNRAS}\
{\elevenbf 212}, 105.\hfill}
\line{Lin, D.N.C., Papaloizou, J.C.B. and Savonije, G.J. 1990, {\elevenit ApJ}\
{\elevenbf 364}, 326.\hfill}
\line{Livio, M. and Spruit H.C. 1991, {\elevenit A\&A}\ {\elevenbf 252},
189.\hfill}
\line{Lynden-Bell, D. 1969, {\elevenit Nature}\ {\elevenbf 223}, 690.\hfill}
\line{Lubow, S.H. 1992, {\elevenit ApJ}\ {\elevenbf 398}, 525.\hfill}
\line{Matsuda, T., Sekino, W., Shima, E., Sawada, K. and Spruit, H.C. 1990,
{\elevenit A\&A}\ {\elevenbf 235},\hfill}
\line{Mauche, C. W., Miller, G. S., Raymond, J. C. and Lamb, F. K. 1990,
in\hfill}
\line{\quad {\elevenit Accretion-Powered Compact Binaries}, Proc. of the 11th
North Am. Workshop\hfill}
\line{\quad on CV and LMXB, ed. C. W. Mauche (Cambridge Univ. Press:
Cambridge), 195.\hfill}
\line{McComas, C.H. and Bretherton, F.P. 1977, {\elevenit J. Geophys. Res.}\
{\elevenbf 83}, 1397.\hfill}
\line{\quad 211.\hfill}
\line{Meyer, F. 1984, {\elevenit A\&A}\ {\elevenbf 131}, 303.\hfill}
\line{Meyer, F. and Meyer-Hofmeister, E. 1984, {\elevenit A\&A}\ {\elevenbf
104}, l10.\hfill}
\line{Mineshige, S. and Osaki, Y. 1983, {\elevenit PASJ}\ {\elevenbf 35},
377.\hfill}
\line{Mineshige, S. and Osaki, Y. 1985, {\elevenit PASJ}\ {\elevenbf 37}, 1.
\hfill}
\line{Mineshige, S. and Wheeler, J.C. 1989, {\elevenit ApJ}\ {\elevenbf 343},
241.\hfill}
\line{Moffatt, H.K. 1978, {\elevenit Magnetic Field Generation in Electrically
Conducting Fluids},\hfill}
\line{\quad (Univ. Press, Cambridge).\hfill}
\line{M\"uller, P., Holloway, G., Henyey, F. and Pomphrey, N. 1986,\hfill}
\line{\quad{\elevenit Rev. Geophys.}\ {\elevenbf 24}, 493.\hfill}
\line{Narayan, R., Goldreich, P. and Goodman, J. 1987, {\elevenit MNRAS}\
{\elevenbf 228}, 1.\hfill}
\line{Noh, H., Vishniac, E.T. and Cochran, W.D. 1992, {\elevenit ApJ}\
{\elevenbf 397}, 347.\hfill}
\line{Paczynski, B. 1976, {\elevenit Comm. Ap.}\ {\elevenbf 6}, 95.\hfill}
\line{Papaloizou, J.C.B., and Pringle, J. 1984, {\elevenit MNRAS}\ {\elevenbf
208}, 721.\hfill}
\line{Papaloizou, J.C.B., and Pringle, J. 1985, {\elevenit MNRAS}\ {\elevenbf
213}, 799.\hfill}
\line{Parker, E.N. 1971, {\elevenit ApJ}\ {\elevenbf 163}, 252.\hfill}
\line{Parker, E.N. 1979, {\elevenit Cosmical Magnetic Fields (their origin and
activity)},\hfill}
\line{\quad (Clarendon Press: Oxford).\hfill}
\line{Phillips, O.M. 1960, {\elevenit J. Fluid Mech.}\ {\elevenbf 9},
193.\hfill}
\line{Phillips, O.M. 1961, {\elevenit J. Fluid Mech.}\ {\elevenbf 11},
145.\hfill}
\line{Pringle, J.E., Rees, J.J., Pacholczyk, A.G. 1973, {\elevenit A\&A}\
{\elevenbf 29}, 179.\hfill}
\line{Pringle, J.E. 1981, {\elevenit Ann. Rev. Astron. Astrophys.}\ {\elevenbf
19} 137.\hfill}
\line{Pudritz, R.E. 1981a, {\elevenit MNRAS}\ {\elevenbf 195}, 881.\hfill}
\line{Pudritz, R.E. 1981b, {\elevenit MNRAS}\ {\elevenbf 195}, 897.\hfill}
\line{Robinson, E.L., Barker, E.S., Cochran, A.L., Cochran, W.D. and Nather,
R.E. 1981,\hfill}
\line{\quad {\elevenit ApJ}\ {\elevenbf 251}, 611.\hfill}
\line{Rozyczka, M. and Spruit, H.C. 1989, {\elevenit Theory of Accretion
Disks}, ed. F. Meyer, W.J.\hfill}
\line{\quad Duschl, J. Frank and E. Meyer-Hofmeister (Kluwer Acad. Pub.:
Dordrecht) 341.\hfill}
\line{Rozyczka, M. and Spruit, H.C. 1992, {\elevenit ApJ}, in press\hfill}
\line{Ruden, S.P., Papaloizou, J.C.B. and Lin, D.N.C. 1988, {\elevenit ApJ}\
{\elevenbf 329}, 739.\hfill}
\line{Ryu, D. and Goodman, J. 1992, {\elevenit ApJ}\ {\elevenbf 388},
438.\hfill}
\line{Sawada, K. and Matsuda, T. 1992, {\elevenit MNRAS}\ {\elevenbf 255},
17p.\hfill}
\line{Sawada, K., Matsuda, T. and Hachisu, I. 1986, {\elevenit MNRAS}\
{\elevenbf 219}, 75.\hfill}
\line{Sawada, K., Matsuda, T., Inoue, T. and Hachisu, I. 1987, {\elevenit
MNRAS}\ {\elevenbf 224}, 307.\hfill}
\line{Shakura, N.I. and Sunyaev, R.A.  1973, {\elevenit A\&A}\ {\elevenbf 22},
471.\hfill}
\line{Shu, F. 1974, {\elevenit A\&A}\ {\elevenbf 33}, 75.\hfill}
\line{Shu, F. 1976, {\elevenit Structure and Evolution of Close Binary
Systems}, ed. P. Eggleton\hfill}
\line{\quad (D. Reidel Pub. Co.: Dordrecht) 253.\hfill}
\line{Shu, F., Tremaine, S., Adams, F.C. and Ruden, S.P. 1990, {\elevenit ApJ}\
{\elevenbf 358}, 495.\hfill}
\line{Smak, J. 1982 , {\elevenit Acta Astr.}\ {\elevenbf 32}, 199.\hfill}
\line{Smak, J. 1984 , {\elevenit Acta Astr.}\ {\elevenbf 34}, 161.\hfill}
\line{Spruit, H.C. 1987, {\elevenit A\&A}\ {\elevenbf 184}, 173.\hfill}
\line{Spruit, H.C., Matsuda, T., Inoue, T., and Sawada, K. 1987, {\elevenit
MNRAS}\ {\elevenbf 229}, 517.\hfill}
\line{Stella, L. and Rosner, R. 1984, {\elevenit ApJ}\ {\elevenbf 277},
312.\hfill}
\line{Tagger, M., Pellat, R. and Coroniti, F.V. 1992, {\elevenit ApJ}\
{\elevenbf 393}, 708.\hfill}
\line{Tagger, M., Henriksen, R.N., Pellat, R. and Sygnet, J.F. 1990, {\elevenit
ApJ}\ {\elevenbf 353}, 654.\hfill}
\line{Tout, C.A. and Pringle, J.E. 1992, {\elevenit MNRAS}\ {\elevenbf 259},
604.\hfill}
\line{Vainshtein, S.I. and Cattaneo, F. 1992, {\elevenit ApJ}\ {\elevenbf 393},
165.\hfill}
\line{Vainshtein, S.I., Parker, E.N. and Rosner, R. 1993, {\elevenit ApJ} in
press.\hfill}
\line{Vainshtein, S.I. and Ruzmaikin, A.A. 1971, {\elevenit Astron. J. (SSSR)}\
{\elevenbf 48}, 902.\hfill}
\line{Vainshtein, S.I. and Ruzmaikin, A.A. 1972, {\elevenit Astron. J. (SSSR)}\
{\elevenbf 49}, 449.\hfill}
\line{Velikhov, E.P. 1959, {\elevenit Soviet JETP}\ {\elevenbf 35}, 1398.
\hfill}
\line{Vila, 1978, {\elevenit ApJ}\ {\elevenbf 223}, 979.\hfill}
\line{Vishniac, E.T. and Diamond, P.H.  1989, {\elevenit ApJ}\ {\elevenbf 347},
435.\hfill}
\line{Vishniac, E.T., Jin, L. and Diamond, P.H. 1990, {\elevenit ApJ}\
{\elevenbf 365}, 552.\hfill}
\line{Vishniac, E.T. and Diamond, P.H. 1992, {\elevenit ApJ}\ {\elevenbf 398},
561.\hfill}
\line{Wood, J. and Mineshige, S. 1989, {\elevenit MNRAS}\ {\elevenbf 241},
259.\hfill}
\line{Zel`dovich, Ya.B. 1956, {\elevenit JETP}\ {\elevenbf 31}, 154.\hfill}
\line{Zel`dovich, Ya.B., Ruzmaikin, A.A. and Sokoloff, D.D. 1983, {\elevenit
Magnetic Fields in}\hfill}
\line{\quad{\elevenit Astrophysics}, (Gordon and Breach Science Publishers: New
York).\hfill}
\line{Zhang, W., Diamond, P.H. and Vishniac, E.T. 1992, submitted to {\elevenit
ApJ}.\hfill}
\vfill
\end